\documentclass[prd,preprint,showpacs,preprintnumbers,nofootinbib,superscriptaddress]{revtex4}

\usepackage{amsmath}
\usepackage{amssymb}
\usepackage{graphicx}
\usepackage{enumerate}
\usepackage{subfigure}

\newcommand{\Tr}{{\rm Tr}} 

\newcommand{\sla}[1]{{/\hskip-0.55em{}#1}{}}

\newcommand{\ff}{\tilde{f}}

\begin{document}



\title{%
Automated Calculation Scheme for $\alpha^n$ Contributions of QED 
to Lepton $g\!-\!2$: 
New Treatment of Infrared Divergence for Diagrams without Lepton Loops}


\author{T.~Aoyama}
\affiliation{Institute of Particle and Nuclear Studies, High Energy Accelerator Research Organization (KEK), Tsukuba, 305-0801, Japan }

\author{M.~Hayakawa}
\affiliation{Department of Physics, Nagoya University, Nagoya, 464-8602, Japan }

\author{T.~Kinoshita}
\affiliation{Laboratory for Elementary-Particle Physics, Cornell University, Ithaca, New York 14853, U.S.A. }

\author{M.~Nio}
\affiliation{Theoretical Physics Laboratory, RIKEN, Wako, 351-0198, Japan }

\begin{abstract}
We have developed an efficient algorithm for the subtraction 
of infrared divergences that arise in the evaluation of QED 
corrections to the anomalous magnetic moment of lepton ($g\!-\!2$). 
By incorporating this new algorithm, we have extended 
the automated code-generating system developed previously 
to deal with diagrams without internal lepton loops 
(called {\it q-type}), 
which produced convergent integrals when applied 
to diagrams that have only ultraviolet-divergent 
subdiagrams of vertex type. 
The new system produces finite integrals for all 
{\it q-type} diagrams, 
including those that contain self-energy subdiagrams 
and thus exhibit infrared-divergent behavior. 
We have thus far verified the system for the sixth- and eighth-order cases. 
We are now evaluating 6354 vertex diagrams of {\it q-type} 
that contribute to the tenth-order lepton $g\!-\!2$. 
\end{abstract}

\pacs{ 13.40.Em, 14.60.Cd, 14.70.Bh, 11.15.Bt, 12.20.Ds }



\maketitle


\section{Introduction}
\label{sec:intro}

The magnetic moment anomaly of electron, 
also called the electron $g\!-\!2$, has been measured with 
steadily increased precision since its discovery in 1947 \cite{Kusch:1947}. 
The latest experiment using Penning trap with cylindrical cavity 
updates the best previous value 
\cite{VanDyck:1987ay}
of the electron $g\!-\!2$ to 
\cite{Odom:2006gg}
\begin{equation}
  a_e = 1~159~652~180.85~(76) \times 10^{-12}
  \quad [0.76\text{ppb}].
\label{eq:ae}
\end{equation}
Here, $a_e = \frac{1}{2}(g-2)$ and the numeral in the parenthesis 
represents uncertainty in the last two digits of the value. 
The uncertainty has been reduced by a factor 5.5 over the previous 
best value. 

Theoretically, the electron $g\!-\!2$ is almost entirely accounted for 
by the QED corrections, and thus it has provided the most 
stringent test of the validity of QED. 
The dominant contribution comes from the interaction between 
photons and electrons alone (referred to as mass-independent term), 
and it is given as a function of the fine structure constant $\alpha$. 
An important by-product of the study of the electron $g\!-\!2$ is 
that a very precise value of $\alpha$ can be obtained from the 
measurement and the theory of $a_e$, 
assuming the validity of QED. 
The value of $\alpha$ thus determined is 
\cite{Gabrielse:2006ggE}
\begin{equation}
  \alpha^{-1}(a_e) = 137.035~999~070~(12)(37)(90)
  \quad [0.71\text{ppb}],
\label{eq:alpha}
\end{equation}
where uncertainties are due to the $\alpha^4$ term, an educated guess 
of the $\alpha^5$ term \cite{CODATA:2002}, 
and the experiment (\ref{eq:ae}). 
The value (\ref{eq:alpha}) is by far the most precise determination 
of $\alpha$ at present\footnote{%
Note that the value given in Eq.~(\ref{eq:alpha}) includes small changes 
in the tenth-order estimate \cite{CODATA:2002} and 
in the hadronic light-by-light contribution 
\cite{CODATA:2002,Melnikov:2003xd}, 
and thus deviates from $\alpha^{-1}$ given in 
Ref.~\cite{Gabrielse:2006gg}.
}. 

As is seen from Eq.~(\ref{eq:alpha}) the measurement uncertainty 
in $\alpha(a_e)$ is only a factor 2.5 larger than that of the theory 
which is mostly due to the unknown $\alpha^5$ term. 
When the experimental data improves further, the $\alpha^5$ term 
will become the largest source of unresolved systematic errors. 
This is why an explicit evaluation of $\alpha^5$ term is urgently needed. 

The mass-independent contribution to the $\alpha^5$ term involves 
12672 Feynman diagrams, 
which can be divided into 32 gauge-invariant sets. 
We classify them into 6 super sets (Set I--VI) according to 
their structures (see Figs.~I--VI of Ref.~\cite{Kinoshita:2005sm}). 
The largest and the most difficult is the Set V, which consists 
of 6354 diagrams that have no lepton loops (we call them 
{\it q-type}) and form a single gauge-invariant set. 
The difficulty of the Set V stems from the fact that 
many of them, besides being gigantic, have very large number of 
ultraviolet (UV) and infrared (IR) divergences. 

In the previous approach employed in the calculation 
of sixth-order and eighth-order diagrams the subtraction terms 
of UV divergence were constructed by a procedure called 
{\it K}-operation \cite{Kinoshita:1990}, which is based on a simple power counting 
at a UV-divergent singularity. 
In order to deal with UV divergences, let us 
regularize each relevant photon propagator 
by introducing the Feynman cut-off 
\begin{equation}
  \frac{1}{k^2}
  \ \longrightarrow\ 
  \frac{1}{k^2 - \lambda^2} - \frac{1}{k^2 - \Lambda^2}
  =
  -\int_{\lambda^2}^{\Lambda^2} \frac{dL}{(k^2-L)^2},
\label{feynmancutoff}
\end{equation}
where $\Lambda$ and $\lambda$ are UV-cutoff and IR-cutoff,
respectively.
Suppose the diagram $\mathcal{G}$ has a UV-divergent 
{\em vertex} subdiagram $\mathcal{S}$. 
Let $M_\mathcal{G}$ be the magnetic moment contribution of $\mathcal{G}$.
Then the operation of 
$\mathbb{K}_\mathcal{S}$ on $M_\mathcal{G}$ creates 
an integral $\mathbb{K}_\mathcal{S} M_\mathcal{G}$ 
which has the same UV cutoff $\Lambda$ as $M_\mathcal{G}$ itself 
and yet factorizable into the product shown on the right-hand side: 
\begin{equation}
	\mathbb{K}_\mathcal{S} M_\mathcal{G}
	=
	L_\mathcal{S}^{\rm UV} M_{\mathcal{G}/\mathcal{S}}, 
\label{Kop1}
\end{equation}
where $L_\mathcal{S}^{\rm UV}$ is the leading UV-divergent part 
(referred to as the most-contracted term) 
of the vertex renormalization constant $L_\mathcal{S}$, 
and $M_{\mathcal{G}/\mathcal{S}}$ is the magnetic moment projection 
for the reduced diagram that is obtained from $\mathcal{G}$ by 
shrinking $\mathcal{S}$ to a point. 
It is important to note that the factorization in Eq.(\ref{Kop1}) 
does not work unless
both sides are well-defined integrals
(made finite by the Feynman cut-off or some other regularization).
Throughout this paper let us assume that
all UV-divergent integrals are regularized by the Feynman cutoff.
Of course the Feynman cutoff is not needed for convergent integrals, 
and the limit $\Lambda\to\infty$ must be taken 
after the renormalization is carried out. 

In Ref.~\cite{Aoyama:2005kf} it was shown that this scheme 
of UV-subtraction 
can be incorporated into an automating algorithm without change. 
In the case of Set V diagrams of tenth-order 
we were able to obtain FORTRAN codes for 2232 vertex diagrams
containing only vertex subdiagrams 
(represented by 135 Ward-Takahashi-summed diagrams
[see Eq.~(\ref{eq:wtsum})])
out of the total of 6354 vertex diagrams 
(represented by 389 Ward-Takahashi-summed diagrams). 
A preliminary numerical evaluation of these integrals 
by VEGAS \cite{vegas} 
shows that UV divergences are removed completely. 

In case where $\mathcal{S}$ is a {\em self-energy} subdiagram 
inserted between consecutive lines $i$ and $j$ of $\mathcal{G}$, 
the action of $\mathbb{K}_\mathcal{S}$ on $M_\mathcal{G}$ yields 
a somewhat more complicated factorization 
\begin{equation}
	\mathbb{K}_\mathcal{S} M_\mathcal{G}
	=
	\delta m_\mathcal{S}^{\rm UV}\, 
         M_{\mathcal{G}/\mathcal{S}\,(i^\star)}
	+
	B_\mathcal{S}^{\rm UV} M_{\mathcal{G}/[\mathcal{S},j]}, 
\label{Kop2}
\end{equation}
where $\delta m_\mathcal{S}^{\rm UV}$ and $B_\mathcal{S}^{\rm UV}$ are 
the leading UV divergent parts of the mass renormalization constant 
$\delta m_\mathcal{S}$ and 
the wave-function renormalization constant 
$B_\mathcal{S}$, respectively. 
The $^\star$ in $M_{\mathcal{G}/\mathcal{S}\,(i^\star)}$ indicates that
it has a two-point vertex between the lines $i$ and $j$. 
$\mathcal{G}/[\mathcal{S},j]$ denotes a diagram 
obtained by shrinking both $\mathcal{S}$ and line $j$ to points. 

This method works as far as subtraction of UV divergence is concerned. 
However, it complicates the handling of IR divergence 
because of the fact that the unrenormalized $M_\mathcal{G}$
as well as its self-mass counter term 
$\delta m_\mathcal{S} M_{\mathcal{G}/\mathcal{S}\,(i^\star)}$ 
have a linear or worse IR divergence. 
In the calculation of the eighth-order case we encountered 
such linear IR divergences in two diagrams\footnote{%
$M_{16}$ and $M_{18}$ share the structure which is obtained by 
inserting two self-energy subdiagrams, 
one is of the second order and the other is of the fourth order, 
into the second-order magnetic moment term.} 
$M_{16}$ and $M_{18}$ 
as well as in some renormalization constants 
\cite{Kinoshita:1981wm,Kinoshita:1990}. 
This problem has been handled by subtracting the linear IR divergence 
by an {\it ad hoc} subtraction term. 

However, such an {\it ad hoc} approach will become very complicated 
in the tenth-order case. 
Furthermore, it will present a severe obstacle to automation. 
Thus, we have developed an alternative approach to subtraction 
of the linear (and worse) IR divergence 
which is entirely systematic and fits well in the scheme 
of automated code generation. 
It generates FORTRAN codes of the renormalized and finite 
amplitudes for all diagrams of the Set V very rapidly, 
which are ready for numerical integration. 

The aim of this paper is to present the new approach for 
constructing the IR subtraction term in an automated manner. 
We also report its implementation. 

The organization of the paper is as follows. 
In Sec.~\ref{sec:amplitude} we briefly summarize 
our formulation of the numerical evaluation of lepton $g\!-\!2$ 
and the scheme of subtractive UV renormalization. 
In Sec.~\ref{sec:ir} we present a new scheme for identifying 
IR divergences that may appear in a complex manner. 
The construction of subtraction integrals for those divergences 
is described in Sec.~\ref{sec:construction}. 
In Sec.~\ref{sec:proc} we show the concrete procedure of 
code generation and its implementation. 
Sec.~\ref{sec:conclusion} is devoted to conclusion and discussion. 
An example of the identification of IR subtraction terms is 
given in Appendix~\ref{sec:example}.


\section{Unrenormalized Amplitude and UV Divergences}
\label{sec:amplitude}

In this section we briefly summarize our formulation for 
evaluating QED contribution to the anomalous magnetic moment 
of leptons by numerical means 
\cite{Cvitanovic:1974uf,Cvitanovic:1974sv,Kinoshita:1990}. 
It involves the construction of amplitudes for Feynman diagrams 
and the renormalization of ultraviolet divergences. 
We note that the substructure called 
{\em forests} 
plays a crucial role in organizing
the UV renormalization and
also accounts for our subtraction scheme of 
infrared divergences discussed in the later sections. 

\subsection{Anomalous Magnetic Moment of Lepton}
\label{sec:amplitude:g-2}

The anomalous magnetic moment $a_e$ is given by the static limit 
of the magnetic form factor that is related to the proper vertex 
part $\Gamma^\nu$. 
We evaluate the QED contribution to $a_e$ in the framework of 
perturbation theory by the series expansion with respect to 
$\alpha/\pi$, where $\alpha$ is the fine structure constant. 

In our formulation 
we employ a relation derived from the Ward-Takahashi identity 
\begin{equation}
	\Lambda^\nu(p,q) 
	\simeq 
	- q^\mu \left[
		\frac{\partial\Lambda_\mu(p,q)}{\partial q_\nu}
	\right]_{q\to0} 
	- \frac{\partial\Sigma(p)}{\partial p_\nu}
\label{eq:wtsum}
\end{equation} 
between the self-energy part $\Sigma(p)$  and 
the sum of vertex parts $\Lambda^\nu(p,q)$ obtained by inserting 
an external vertex in the lepton lines of $\Sigma$ 
in all possible ways. 
Here, the momentum of the incoming lepton $p-\frac{1}{2}q$ and 
that of the outgoing lepton $p+\frac{1}{2}q$ are on the mass shell 
so that $p$ and $q$ satisfy 
$p^2=m^2-\frac{1}{4}q^2$ and $p \cdot q = 0$. 
Eq. (\ref{eq:wtsum}) enables us to turn 
the evaluation of a set of vertex diagrams 
into the evaluation of a single self-energy-like diagram. 
This approach reduces the number of independent diagrams substantially. 

The magnetic moment anomaly of a $2n$th-order diagram $\mathcal{G}$ 
is given by an integral over loop momenta of a product of vertices 
and propagators of lepton and photon. 
It is turned into a parametric integral over Feynman parameters 
$z_i$ by the Feynman integration formula. 
Carrying out the momentum integration analytically, 
we can express the resulting amplitude in a concise form: 
\begin{multline}
  M_\mathcal{G}^{(2n)} 
  = 
  \left(-\frac{1}{4}\right)^{n} (n-1)! 
  \int (dz)_\mathcal{G}
  \left[
    \frac{1}{n-1}\left(
    \frac{E_0 + C_0}{U^2 V^{n-1}} + \frac{E_1 + C_1}{U^3 V^{n-2}} + \cdots
    \right)
\right. \\
    +
\left .
    \left(
    \frac{N_0 + Z_0}{U^2 V^{n}} + \frac{N_1 + Z_1}{U^3 V^{n-1}} + \cdots
    \right)
  \right],
\label{parametricint}
\end{multline}
where $(dz)_\mathcal{G} = \prod d z_i \delta(1-\sum_i z_i)$. 
The factor  $(\alpha/\pi)^n$ is omitted for simplicity.
It is implicitly assumed that the Feynman cutoff (\ref{feynmancutoff})
is introduced whenever it is necessary.

The quantities $C_k$, $E_k$, $N_k$, and $Z_k$ are polynomials 
of symbols called building blocks $B_{ij}$, $A_i$, and $C_{ij}$ 
\cite{Cvitanovic:1974uf}. 
The symbols $B_{ij}$ and $U$ are homogeneous polynomials of Feynman 
parameters, related to the flow of loop momenta on the diagram. 
The symbol $A_i$ is called scalar current that is associated 
with the flow of the external momentum $p$. 
It is constructed from $B_{ij}$, $U$, and $\{z_i\}$. 
The symbol $C_{ij}$ is also constructed from $B_{ij}$, $U$, and $\{z_i\}$. 
The symbol $V$ in the denominator is a function defined by 
\begin{equation}
  V = \sum_i z_i - G, \qquad G = \sum_i z_i A_i,
\label{denominatorV}
\end{equation}
where the summation is over the lepton lines only 
and the rest mass of the lepton is set to $m=1$ for simplicity. 


\subsection{Subtraction of UV Divergences}
\label{sec:amplitude:uv}

The amplitude constructed above is UV-divergent 
except for the case $n=1$. 
To achieve pointwise cancellation of these divergences 
we adopt here the subtractive renormalization. 

The UV divergence arises when 
some of the loop momenta go to infinity. 
In the Feynman parametric space it 
corresponds to a particular regime 
in which the sum of relevant Feynman parameters tends to zero. 
We prepare the subtraction term in the form of an integral 
such that it cancels the singularity of the original integrand 
at a singular point of the integration domain. 
These subtraction integrals are constructed 
from the original integrand by the {\it K}-operation. 
As was noted already they factorize exactly into a product 
(or a sum of products) of lower-order magnetic projection 
and a UV-divergent part of the renormalization constant. 
(See Eq. (\ref{Kop1}) and Eq. (\ref{Kop2}).) 

In the prescription described above, 
we only subtracted away a part of 
renormalization constants. 
This is because the integral becomes highly intractable 
if those renormalization constants are treated as a whole. 
To carry out the standard on-shell renormalization, 
we have to take account of the difference, 
{\it e.g.} $L_\mathcal{S} - L_\mathcal{S}^{\rm UV}$ in Eq.~(\ref{Kop1}).
Similarly for the self-mass term and the wave-function renormalization 
term. 
We adopt two-step renormalization, and adjust the differences afterward 
by collecting all contributions over the diagrams. 
We call it the residual renormalization step. 


\subsection{Forests}
\label{sec:amplitude:forests}

A diagram may have complicated divergence structure due to 
nested singularities. 
The whole structure of UV divergences is managed by 
Zimmermann's forest formula. 
A forest is a set of divergent subdiagrams 
and each forest corresponds to a particular emergence of divergence. 

First we introduce an inclusion relation of subdiagrams as follows. 
If two subdiagrams share neither vertex nor line, they are 
called {\it disjoint}. 
If a subdiagram $\mathcal{S}$ is included 
in the other subdiagram $\mathcal{S}^\prime$ in the 
sense that all vertices and lines of $\mathcal{S}$ are also 
the elements of $\mathcal{S}^\prime$, 
$\mathcal{S}$ is {\it included} in $\mathcal{S}^\prime$. 
When $\mathcal{S}$ and $\mathcal{S}^\prime$ have common vertices 
and/or lines but one is not included in the other, 
they are {\it overlapping}. 
In this case there are some vertices and lines of $\mathcal{S}$ 
that are not the elements of $\mathcal{S}^\prime$. 

A forest $f$ is defined as a set of subdiagrams of which any two 
elements are not overlapping. 
It is called normal forest when the diagram $\mathcal{G}$ itself 
is not an element of the forest. 
The whole set of (normal) forests is found by generating all 
possible combinations of subdiagrams and disregarding any 
combinations in which a pair of elements are overlapping. 

Noting that the {\it K}-operation can be defined for each $\mathcal{S}$, 
we can define the UV-finite amplitude $\underline{M}_\mathcal{G}$ 
by the formula 
\begin{equation}
	\underline{M}_\mathcal{G} 
	= 
	M_\mathcal{G}
	+ 
	\sum_{f\in\mathfrak{F}}\,\prod_{\mathcal{S}\in f} 
	\left(-\mathbb{K}_\mathcal{S}\right)\,M_\mathcal{G}, 
\label{forestformula}
\end{equation} 
where the summation is taken over the set  
$\mathfrak{F}$ 
of the normal forests 
of the diagram $\mathcal{G}$, 
and the product means the successive application of {\it K}-operations. 


\section{IR Divergences}
\label{sec:ir}

The magnetic form factor is free from UV- and IR-divergences 
once it is fully renormalized. 
However, individual diagrams suffer from IR divergences 
which cancel out only after all diagrams are combined. 

The root cause of IR divergence is the vanishing of the denominator 
of the photon propagator $1/k^2$ in the limit $k \rightarrow 0$. 
This is, however, not the sufficient condition since it gives 
a finite result on integration over the 4-dimensional momentum $k$. 
In order that it becomes divergent, it must be enhanced 
by vanishing of the denominators of at least two lepton propagators 
due to some kinematical constraints. 
Typically, this happens when the momentum of each of 
these lepton propagators is constrained by sharing a three-point 
vertex with the soft photon and an external on-shell lepton line. 
When the external momentum $p$ is constrained 
by the on-shell condition $p^2 = m^2$, 
the lepton propagator in question behaves as 
\begin{equation}
  \frac{1}{(p+k)^2-m^2} = \frac{1}{2p \cdot k+k^2} 
  \sim \frac{1}{2 p \cdot k} \,  
\end{equation}
for $k \rightarrow 0$. 
These lepton propagators will be called ``enhancers''. 
The logarithmic IR divergence 
takes place when the $k$-integration is carried out 
and the soft photon singularity is assisted by two enhancers. 
When the vertex Feynman diagram $\mathcal{G}(k)$ in question 
has a self-energy subdiagram, 
we find three enhancers due to the kinematical constraint 
of the two-point vertex so that we find the IR divergence 
to be linear. 
The IR divergence becomes even severer when the diagram 
$\mathcal{G}(k)$ has more than one self-energy subdiagram 
which effectively bring in a number of two-point vertices. 

To handle the IR divergences, we adopt again a subtractive approach 
in which an integral of IR subtraction terms is constructed 
in such a way that it cancels out the IR-divergence 
of the integral $M_\mathcal{G}^{(2n)}$ 
of Eq. (\ref{parametricint}) point-by-point 
in the Feynman parameter space. 

First we briefly summarize a scheme called {\it I}-operation 
\cite{Kinoshita:1990} 
which has been developed and employed for the calculation of 
sixth- and eighth-order calculations. 

In the Feynman-parametric representation, 
the IR divergence is caused by vanishing of the 
denominator function $V$ of Eq.(\ref{denominatorV}) 
in the corner of the integration domain characterized by 
\begin{align}
  z_i &= {\cal O} (\delta)
  &&\text{if $i$ is an lepton line in $\mathcal{R}$},  \nonumber \\
  z_i &= {\cal O} (1)
  &&\text{if $i$ is a photon line in $\mathcal{R}$},     \nonumber \\
  z_i &= {\cal O} (\epsilon),\quad \epsilon\sim\delta^2
  &&\text{if $i \in \mathcal{S}$},
\label{irlimit}
\end{align}
where $\mathcal{R} = \mathcal{G}/\mathcal{S}$. 
In this limit the denominator $V$ vanishes as ${\cal O} (\delta^2)$. 
(The last condition is actually an artifact of the 
condition $\sum_i z_i = 1$, which can be readily lifted.) 

If two lepton propagators participate in the enhancement, 
we obtain a {\it logarithmic} IR divergence. 
In this case we can construct an IR subtraction term 
by a simple power counting rule and an $I$-operation similar 
to the $K$-operation of the UV divergent case. 
For the subdiagram $\mathcal{R}= \mathcal{G}/\mathcal{S}$ 
the {\it I}-operation $I_\mathcal{R}$ is defined as follows: 
\begin{enumerate}[(a)]
\item In the limit (\ref{irlimit}) keep only terms with lowest 
power of $\epsilon$ and $\delta$ in $U, B_{ij}, A_i$. 
\item Make the following replacements: 
\begin{equation}
  U \rightarrow U_\mathcal{S} U_\mathcal{R},
  \qquad
  V \rightarrow V_\mathcal{S} + V_\mathcal{R},
  \qquad
  F \rightarrow F_0 [L_\mathcal{R}] F_\mathcal{S},
\end{equation}
where $F_0 [L_\mathcal{R}]$ is the non-contracting term of the 
vertex renormalization constant defined on $\mathcal{R}$, 
and $F_\mathcal{S}$ is the product of $\gamma$ matrices and 
$D_i^\mu$ operators for the diagram $\mathcal{S}$. 
(See Ref.~\cite{Kinoshita:1990} for definitions.) 
\end{enumerate}
This procedure creates an integral defined on the parametric space
of $M_\mathcal{G}$. By construction it factorizes as
\begin{equation}
  L_{\mathcal{R}(k)}[F_0]\,M_\mathcal{S}
  + 
  M_{\mathcal{R}^\star}[I]\,\Delta\delta\tilde{m}_{\mathcal{S}}
  \,.
\label{eq:qedsub}
\end{equation}
For precise definitions of quantities quoted above 
see Ref.~\cite{Kinoshita:1990}. 

In the following, we see that 
for the lepton $g\!-\!2$ of the {\it q-type} diagrams, 
there are two kinds of sources of enhancement. 
These two types of divergences may occur simultaneously 
in a complicated manner. 
We will introduce a new scheme modeled on the previous approach 
for identifying the emergence of the divergences. 
In this scheme the identification relies on examining the shape 
of the diagram that allows diagrammatic treatment systematically 
without the need for close examination of the integrand, 
and it is also suited for automated treatment. 


\subsection{IR Divergence Caused by Residual Self-mass }
\label{sec:ir:r-sub}

One type of IR divergence appears as a consequence of our particular 
treatment of the UV divergences by means of $K$-operation. 
Suppose a diagram $\mathcal{G}$ has a self-energy subdiagram 
$\mathcal{S}$. 
As is readily seen from the analysis of Feynman diagrams, 
this divergence is not the source of real problem since it 
must be canceled exactly by the mass-renormalization counterterm 
${\delta m}_{\mathcal{S}}\,M_{\mathcal{G}/\mathcal{S}\,(i^\star)}$, 
where $\delta m_{\mathcal{S}}$ is the (UV-divergent) self-mass 
associated with the subdiagram $\mathcal{S}$ defined on the mass shell. 
The reduced magnetic moment amplitude 
$M_{\mathcal{G}/\mathcal{S}\,(i^\star)}$ is the one 
that has a linear IR divergence. 
As a consequence 
\begin{equation}
  M_{\mathcal{G}} -
  {\delta m}_{\mathcal{S}}\,M_{\mathcal{G}/\mathcal{S}\,(i^\star)}
\end{equation}
is free from linear IR divergence. 
Although this cancellation is analytically valid, however, 
it is not a pointwise cancellation in the domain of $M_{\mathcal{G}}$. 
Our problem is thus to translate the second term 
into a form which is defined in the same domain as that of 
$M_{\mathcal{G}}$ and cancels the IR divergence of 
$M_{\mathcal{G}}$ point-by-point. 

Now, as was noted in Eq. (\ref{Kop2}) the {\it K}-operation 
for the subdiagram $\mathcal{S}$ acting on $M_\mathcal{G}$ creates 
\begin{equation}
  \mathbb{K}_\mathcal{S} M_\mathcal{G} 
  = 
  \delta m_{\mathcal{S}}^{\rm UV}\,
  M_{\mathcal{G}/\mathcal{S}\,(i^\star)}
  + 
  B_{\mathcal{S}}^{\rm UV}\,
  M_{\mathcal{G}/[\mathcal{S},j]}
  \,. 
\label{eq:k-op-se}
\end{equation}
If we find an integral what causes pointwise cancellation of 
the linear IR divergence in the domain of $M_{\mathcal{G}}$ 
and also produces the factorization as 
\begin{equation}
  \widetilde{\delta m}_{\mathcal{S}} 
  M_{\mathcal{G}/\mathcal{S}\,(i^\star)}
  \,
\label{eq:r-op-tilde}
\end{equation}
where 
\begin{equation}
  \widetilde{\delta m}_{\mathcal{S}} 
  \equiv 
  \delta m_{\mathcal{S}} - {\delta m}_{\mathcal{S}}^{\rm UV}
  \,,
\label{eq:dm-tilde}
\end{equation}
then from Eqs.~(\ref{eq:k-op-se}) and (\ref{eq:r-op-tilde}) 
we would have 
\begin{equation}
  \mathbb{K}_\mathcal{S} M_\mathcal{G} 
  + \widetilde{\delta m}_{\mathcal{S}} M_{\mathcal{G}/\mathcal{S}\,(i^\star)}
  = 
  \delta m_{\mathcal{S}}\,
  M_{\mathcal{G}/\mathcal{S}\,(i^\star)}
  + 
  B_{\mathcal{S}}^{\rm UV}\,
  M_{\mathcal{G}/[\mathcal{S},j]}
  \,. 
\label{eq:k-op-se-2}
\end{equation}

If we schematically introduce an operator $\mathbb{R}_{\mathcal{S}}$ 
that produces the integral of Eq.~(\ref{eq:r-op-tilde}) as 
\begin{equation}
  \mathbb{R}_{\mathcal{S}} M_{\mathcal{G}} 
  \equiv
  \widetilde{\delta m}_{\mathcal{S}} 
  M_{\mathcal{G}/\mathcal{S}\,(i^\star)}
  \,,
\label{eq:r-op-schematic}
\end{equation}
Eq.~(\ref{eq:k-op-se-2}) would then be written as 
\begin{equation}
  (\mathbb{K}_{\mathcal{S}}+\mathbb{R}_{\mathcal{S}}) M_{\mathcal{G}} 
  =
  \delta m_{\mathcal{S}}\,
  M_{\mathcal{G}/\mathcal{S}\,(i^\star)}
  + 
  B_{\mathcal{S}}^{\rm UV}\,
  M_{\mathcal{G}/[\mathcal{S},j]}
  \,.
\end{equation}
It turns out that it is not difficult to construct such an integral. 
Furthermore, it can be readily incorporated in our automation algorithm. 
We call this subtraction scheme as the residual self-mass subtraction, 
or ``{\it R}-subtraction'' operation. 

\subsection{Modified {\it I}-subtraction operation}
\label{sec:ir:i-sub}

\begin{figure}
\caption{%
A vertex diagram $\mathcal{G}(k)$ and a subdiagram $\mathcal{S}(k)$, 
The reduced vertex diagram $\mathcal{R}(k)$ with a vertex 
$M_{\mathcal{S}(k)}$.}
\begin{center}
\subfigure[][]{%
	\includegraphics[scale=1.0]{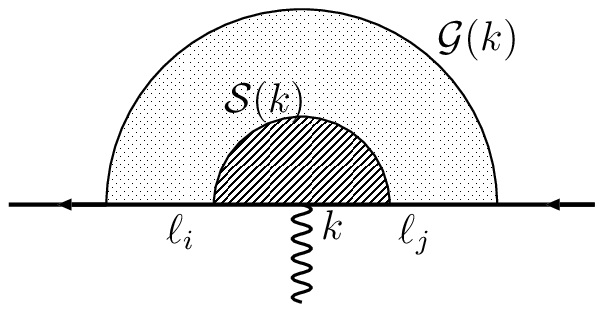}
	\label{fig:v_gs}
}
\hskip1em
\subfigure[][]{%
	\includegraphics[scale=1.0]{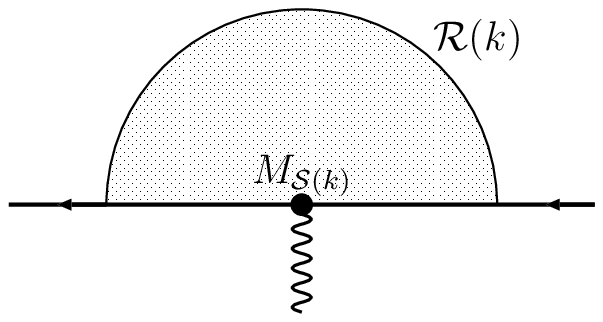}
	\label{fig:v_ae}
}
\end{center}
\label{fig:v_gs_ae}
\end{figure}

After the linear IR divergences are disposed by the 
{\it K}-operation and {\it R}-subtraction operation 
we are still left with logarithmic IR divergences. 
To treat these divergences let us consider 
a vertex diagram $\mathcal{G}(k)$ which has 
a subdiagram $\mathcal{S}(k)$. 
Here $k$ refers to an external photon vertex 
attached to a lepton line $\ell_k$ of $\mathcal{S}$. 
The reduced diagram 
$\mathcal{R} \equiv \mathcal{G}(k)/\mathcal{S}(k)$ 
is connected to $\mathcal{S}(k)$ by 
lepton lines $\ell_i$ and $\ell_j$. 
(See Fig.~\ref{fig:v_gs_ae}\subref{fig:v_gs}.) 

This diagram exhibits an IR-divergent behavior 
when the momenta of (all or some) photons in $\mathcal{R}$ go to zero, 
accompanied by the enhancers $\ell_i$ and $\ell_j$. 
The substructure $\mathcal{S}(k)$ to which these enhancers 
are attached can be considered as a magnetic moment of the lower order. 
Thus the amplitude in this limit becomes that of the diagram 
$\mathcal{R}(k)$ obtained by replacing $\mathcal{S}(k)$ 
by a vertex that is weighted by $M_{\mathcal{S}(k)}$, 
as shown in Fig.~\ref{fig:v_gs_ae}\subref{fig:v_ae}. 
$\mathcal{R}(k)$ develops 
a logarithmic IR divergence as is easily verified 
by power counting. 

Since we are dealing with the Ward-Takahashi-summed 
diagram defined by Eq.~(\ref{eq:wtsum}), 
we have to consider the sum of contributions of vertex diagrams 
$\mathcal{S}(k)$ which are obtained by inserting an external 
vertex $k$ to the self-energy-like diagram $\mathcal{S}$ in 
every possible way. 
In the calculation of the diagram $\mathcal{G}$, 
the IR singularity associated with the substructure $\mathcal{S}$ 
has a form that consists of contributions from 
the vertex diagram $\mathcal{R}(k)$ 
and the magnetic projection of the diagram $\mathcal{S}$. 
The IR singularity is contained 
in the vertex renormalization constant $L_{\mathcal{R}(k)}$. 

For the explicit expression of the IR subtraction term that cancels 
the singularity in the above, 
we consider the term 
\begin{equation}
  \widetilde{L}_{\mathcal{G}/\mathcal{S}(k)} M_{\mathcal{S}}, 
\label{eq:isub}
\end{equation}
where $\widetilde{L}$ is the residual part of the vertex renormalization 
constant 
\begin{equation}
  \widetilde{L}_{\mathcal{G}/\mathcal{S}(k)}
  \equiv
  L_{\mathcal{G}/\mathcal{S}(k)}
  -
  {L}_{\mathcal{G}/\mathcal{S}(k)}^{\rm UV}
  \,.
\end{equation}
We construct an integral that corresponds to Eq.~(\ref{eq:isub}) 
in the domain of $M_{\mathcal{G}}$. 
We call this subtraction scheme as the ``{\it I}-subtraction'' operation. 

Note that the 
subtraction scheme for IR divergence adopted here is different 
from that of Ref.\cite{Kinoshita:1990}. 
We now choose as the IR subtraction term 
the part of $L_{\mathcal{G}/\mathcal{S}(k)}$ 
which excludes  the overall UV divergent part but 
includes a finite part $\Delta L_{\mathcal{G}/\mathcal{S}(k)}$ 
left out in the previous method. 
This simplifies the handling of the IR problem considerably. 

The magnetic moment part $M_\mathcal{S}$ may have UV and IR divergences, 
which also have to be subtracted. 
The UV subdivergences of $M_\mathcal{S}$ 
will be treated in a similar manner to the 
UV subdivergences of $L$. 
The IR divergence that is related to the nested singularity 
is discussed in the next subsection. 

\subsection{Nested Singularity}
\label{sec:ir:nested}

We have introduced the operations for identifying the subtraction 
terms called {\it I}- and {\it R}-subtraction operations 
corresponding to two types of divergences. 
There may appear more than one source of IR divergences that 
lead to complicated divergence structure. 
They are treated by combinations of {\it I}-/{\it R}-subtraction 
operations conducted by {\it annotated forests} described later. 

By generalizing Eq.~(\ref{eq:r-op-schematic}) 
the {\it R}-subtraction operation $\mathbb{R}_{\mathcal{S}}$ 
for a self-energy subdiagram $\mathcal{S}$ 
is defined by 
\begin{equation}
  \mathbb{R}_\mathcal{S}
  \underline{M}_{\mathcal{G}}
  \equiv
  {\delta m}_{\mathcal{S}}^{\rm R}\,
  \underline{M}_{\mathcal{G}/\mathcal{S}\,(i^\star)}
  \,,
\label{eq:def:r-op}
\end{equation}
where the operator $\mathbb{R}_{\mathcal{S}}$ acts 
on the UV-finite amplitude $\underline{M}_{\mathcal{G}}$ 
of the self-energy-like diagram $\mathcal{G}$ 
and generates a product of 
a residual part of mass renormalization constant 
${\delta m}_{\mathcal{S}}^{\rm R}$ defined by Eq.~(\ref{eq:def:tilde-dm}) 
and 
a UV-finite lower-order amplitude 
$\underline{M}_{\mathcal{G}/\mathcal{S}\,(i^\star)}$. 
The reduced diagram $\mathcal{G}/\mathcal{S}\,(i^\star)$ is obtained from 
$\mathcal{G}$ by shrinking $\mathcal{S}$ to a point. 
The term ${\delta m}_{\mathcal{S}}^{\rm R}$ is defined as 
a residual part of the mass renormalization constant 
$\delta m_{\mathcal{S}}$ 
defined on the mass shell 
by subtracting the leading UV divergence 
(referred to as the most-contracted term) 
$\delta m_{\mathcal{S}}^{\rm UV}$ and the subdivergences as 
\begin{equation}
  {\delta m}_{\mathcal{S}}^{\rm R}
  \equiv
  \delta m_{\mathcal{S}}
  - 
  \delta m_{\mathcal{S}}^{\rm UV}
  +
  \sum_{f}\,\prod_{\mathcal{S}^\prime\in f}
  \left(-\mathbb{K}_{\mathcal{S}^\prime}\right)\,
  \widetilde{\delta m}_{\mathcal{S}}
  \,.
\label{eq:def:tilde-dm}
\end{equation}
$\widetilde{\delta m}_{\mathcal{S}}\equiv{\delta m}_{\mathcal{S}}-{\delta m}^{\rm UV}_{\mathcal{S}}$ 
may also have extra UV divergences due to its substructures that 
are subtracted by the {\it K}-operations as above. 
Here, the summation is taken over the normal forests of the subdiagram 
$\mathcal{S}$. 

The {\it I}-subtraction operation $\mathbb{I}_{\mathcal{S}}$ 
for a self-energy subdiagram $\mathcal{S}$ 
is defined by 
\begin{equation}
  \mathbb{I}_\mathcal{S}
  \underline{M}_{\mathcal{G}}
  \equiv
  {L}_{\mathcal{G}/\mathcal{S}(k)}^{\rm R}\,
  \underline{M}_{\mathcal{S}}
  \,,
\label{eq:def:i-op}
\end{equation}
where the operator $\mathbb{I}_{\mathcal{S}}$ acts 
on the UV-finite amplitude $\underline{M}_{\mathcal{G}}$ 
of the self-energy-like diagram $\mathcal{G}$ 
and generates a product of 
a residual part of vertex renormalization constant 
${L}_{\mathcal{G}/\mathcal{S}(k)}^{\rm R}$ 
defined by Eq.~(\ref{eq:def:tilde-l})
and 
a UV-finite lower-order amplitude $\underline{M}_{\mathcal{S}}$. 
Here, the diagram $\mathcal{R}(k)\equiv\mathcal{G}/\mathcal{S}(k)$ 
is given from $\mathcal{G}$ by replacing $\mathcal{S}$ 
by an external photon vertex labelled by $k$.
The term ${L}_{\mathcal{R}(k)}^{\rm R}$ is defined as 
a residual part of the vertex renormalization constant 
$L_{\mathcal{R}(k)}$ 
defined on the mass shell 
by subtracting the leading UV divergence 
(referred to as the most-contracted term) 
$L_{\mathcal{R}(k)}^{\rm UV}$ and the subdivergences as 
\begin{equation}
  {L}_{\mathcal{R}(k)}^{\rm R}
  \equiv
  L_{\mathcal{R}(k)}
  - 
  L_{\mathcal{R}(k)}^{\rm UV}
  +
  \sum_f\,\prod_{\mathcal{S}^\prime\in f}
  \left(-\mathbb{K}_{\mathcal{S}^\prime}\right)\,
  \widetilde{L}_{\mathcal{R}(k)}
  \,.
\label{eq:def:tilde-l}
\end{equation}
$\widetilde{L}_{\mathcal{R}(k)}\equiv{L}_{\mathcal{R}(k)}-{L}_{\mathcal{R}(k)}^{\rm UV}$ 
may also have extra UV divergences due to its substructures that 
have been subtracted by the {\it K}-operations. 
The summation in Eq.~(\ref{eq:def:tilde-l}) is taken 
over the normal forests of the diagram $\mathcal{R}(k)$. 

For a diagram $\mathcal{G}$ containing
a single self-energy subdiagram $\mathcal{S}$, the associated 
IR divergences are treated by those two types of operations, and 
the IR-finite amplitude is thus given by 
\begin{equation}
  \Delta M_{\mathcal{G}}
  = \underline{M}_{\mathcal{G}}
  - \mathbb{I}_{\mathcal{S}} \underline{M}_{\mathcal{G}}
  - \mathbb{R}_{\mathcal{S}} \underline{M}_{\mathcal{G}}
  \,. 
\end{equation}
When the diagram has more than one such self-energy subdiagram, 
the IR divergences due to all those subdiagrams have to be subtracted away. 
The finite amplitude free from both IR and UV divergences are provided 
schematically by 
\begin{equation}
  \Delta M_{\mathcal{G}}
  =
  \prod_{\mathcal{S}} \left(
  1 - \mathbb{I}_{\mathcal{S}} - \mathbb{R}_{\mathcal{S}}
  \right)
  \underline{M}_{\mathcal{G}}
  \,,
\label{eq:subtr:ir}
\end{equation}
where 
the product is taken over all self-energy subdiagrams 
of the diagram $\mathcal{G}$. 

By expanding the product in Eq.~(\ref{eq:subtr:ir}), we are 
lead to a forest-like structure which is analogous to the 
renormalization of UV divergences. 
In this case, a forest consists of only self-energy subdiagrams, 
and each subdiagram is assigned a distinction of {\it R}-subtraction 
or {\it I}-subtraction. 
We call such a forest ``annotated forest''. 
Eq.~(\ref{eq:subtr:ir}) is thus turned into a sum over all 
annotated forests $\ff$ as 
\begin{equation}
  \sum_{\ff}\,
  \left(-\mathbb{I}_{\mathcal{S}_i}\right)\cdots
  \left(-\mathbb{R}_{\mathcal{S}_j}\right)\cdots
  \underline{M}_{\mathcal{G}}
  \,,
\label{eq:subtr:aforest}
\end{equation}
where $\mathcal{S}_i,\dots$ and $\mathcal{S}_j,\dots$ are elements 
of the annotated forest ${\ff}$ that are assigned to 
{\it I}-subtraction operations and {\it R}-subtraction operations, 
respectively. 
Those operators act on the amplitude $\underline{M}_{\mathcal{G}}$ 
successively in an order given below. 

By construction of {\it I}- and {\it R}-operators, there are 
some annotations that are not accepted. 
To see which annotations are allowed, it is sufficient to 
examine the cases in which two self-energy subdiagrams are concerned. 
The inclusion relation of two self-energy subdiagrams is disjoint 
or inclusive. 
Therefore, the possible patterns are exhausted by the cases 
in which 
both subdiagrams are assigned to {\it I}-operators, 
both subdiagrams are assigned to {\it R}-operators, 
or one subdiagram is assigned to {\it I}-operator and the other 
to {\it R}-operator, for two self-energy subdiagrams which 
are disjoint or inclusive. 
We elucidate on those cases separately as follows. 

\begin{figure}
  \caption{%
Successive operations of two {\it I}-subtractions for subdiagrams 
$\mathcal{S}_1$ and $\mathcal{S}_2$ that satisfy 
$\mathcal{S}_1 \supset \mathcal{S}_2$. 
They yield a product of component terms shown on the right-hand side. }
  \begin{center}
    \includegraphics[scale=0.8]{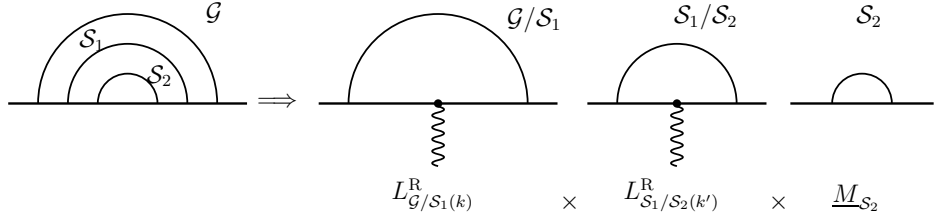}
  \end{center}
  \label{fig:ii_nest}
\end{figure}

First we consider the case which involves two {\it I}-subtractions. 
Recall that the action of operator $\mathbb{I}_{{\mathcal{S}_1}}$ 
is given in Eq.~(\ref{eq:def:i-op}). 
It is found that 
an extra operation of operator $\mathbb{I}_{{\mathcal{S}_2}}$ 
for the subdiagram ${\mathcal{S}_2}$ must have its domain in 
$M_{\mathcal{S}_1}$. 
Therefore the subdiagrams have to respect the relation 
${\mathcal{S}_1}\supset{\mathcal{S}_2}$. 
Otherwise, the operation of $\mathbb{I}_{{\mathcal{S}_2}}$ 
turns to zero. 
It leads to a rule that 
subdiagrams for {\it I}-operators 
must satisfy that one subdiagram is included in the other, 
and a pair of {\it I}-operators for disjoint subdiagrams 
${\mathcal{S}_1}\cap{\mathcal{S}_2}=\emptyset$ 
is not allowed in the annotation. 
The ordering of {\it I}-subtraction operators is given by 
$\mathbb{I}_{{\mathcal{S}_2}}\mathbb{I}_{{\mathcal{S}_1}}$ for 
${\mathcal{S}_1}\supset{\mathcal{S}_2}$. 
(The outer subdiagrams are operated first.) 
\begin{equation}
  \mathbb{I}_{{\mathcal{S}_2}}
  \mathbb{I}_{{\mathcal{S}_1}}
  \underline{M}_{\mathcal{G}}
  =
  {L}_{\mathcal{G}/{\mathcal{S}_1}(k)}^{\rm R}
  {L}_{{\mathcal{S}_1}/{\mathcal{S}_2}(k^\prime)}^{\rm R}
  \underline{M}_{{\mathcal{S}_2}}
  \qquad
  \text{for ${\mathcal{S}_1}\supset{\mathcal{S}_2}$. (See Fig.~\ref{fig:ii_nest}.)}
\end{equation}

\begin{figure}
  \caption{%
Successive operations of two {\it R}-subtractions for subdiagrams 
$\mathcal{S}_1$ and $\mathcal{S}_2$ that satisfy 
$\mathcal{S}_1 \subset \mathcal{S}_2$. }
  \begin{center}
    \includegraphics[scale=0.8]{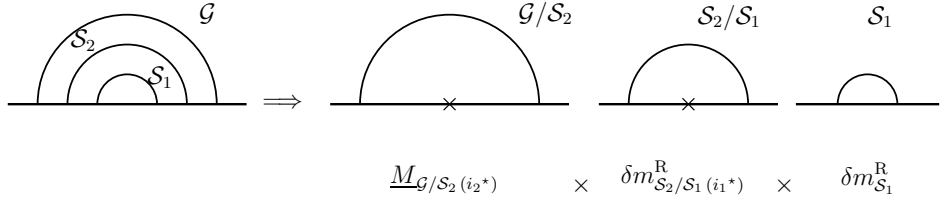}
  \end{center}
  \label{fig:rr_nest}
\end{figure}

\begin{figure}
  \caption{%
Successive operations of two {\it R}-subtractions for subdiagrams 
$\mathcal{S}_1$ and $\mathcal{S}_2$ when they are disjoint, 
$\mathcal{S}_1 \cap \mathcal{S}_2 = \emptyset$. }
  \begin{center}
    \includegraphics[scale=0.8]{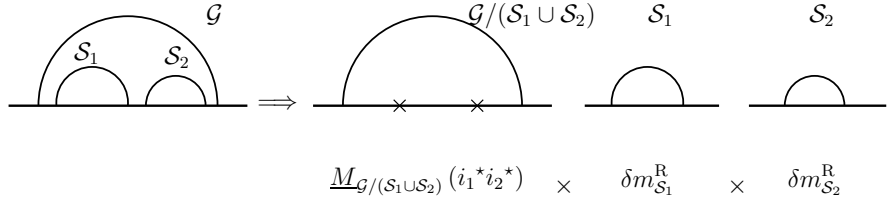}
  \end{center}
  \label{fig:rr_disj}
\end{figure}

The {\it R}-operator for a subdiagram ${\mathcal{S}_1}$ acts as 
Eq.~(\ref{eq:def:r-op}). 
If an extra operator of $\mathbb{R}_{{\mathcal{S}_2}}$ is 
applied, ${\mathcal{S}_2}$ and the reduced diagram 
$\mathcal{G}/{\mathcal{S}_1}\,(i_1^\star)$ 
must share some parts. 
Thus ${\mathcal{S}_1}$ is included in ${\mathcal{S}_2}$, 
${\mathcal{S}_1}\subset{\mathcal{S}_2}$, 
or 
${\mathcal{S}_1}$ and ${\mathcal{S}_2}$ are disjoint. 
The ordering of two {\it R}-operators is given by\footnote{%
The operation of $\mathbb{R}_{{\mathcal{S}_2}}$ 
in this case is actually recognized as 
$\mathbb{R}_{{\mathcal{S}_2}/{\mathcal{S}_1}}$. 
} 
$\mathbb{R}_{{\mathcal{S}_2}}\,\mathbb{R}_{{\mathcal{S}_1}}$ 
for ${\mathcal{S}_1}\subset{\mathcal{S}_2}$. 
(The inner subdiagrams are operated first.) 
\begin{equation}
  \mathbb{R}_{{\mathcal{S}_2}}
  \mathbb{R}_{{\mathcal{S}_1}}
  \underline{M}_{\mathcal{G}}
  =
  \underline{M}_{\mathcal{G}/{\mathcal{S}_2}\,(i_2^{\star})}
  {\delta m}_{{\mathcal{S}_2}/{\mathcal{S}_1}\,(i_1^\star)}^{\rm R} 
  {\delta m}_{{\mathcal{S}_1}}^{\rm R} 
  \qquad
  \text{for ${\mathcal{S}_1}\subset{\mathcal{S}_2}$. 
    (See Fig.~\ref{fig:rr_nest}.)}
\end{equation}
For disjoint subdiagrams, the ordering is indifferent, 
\begin{equation}
  \mathbb{R}_{{\mathcal{S}_2}}
  \mathbb{R}_{{\mathcal{S}_1}}
  \underline{M}_{\mathcal{G}}
  =
  \mathbb{R}_{{\mathcal{S}_1}}
  \mathbb{R}_{{\mathcal{S}_2}}
  \underline{M}_{\mathcal{G}}
  =
  \underline{M}_{\mathcal{G}/({\mathcal{S}_1}\cup{\mathcal{S}_2})\,(i_1^\star i_2^{\star})}
  {\delta m}_{{\mathcal{S}_2}}^{\rm R}
  {\delta m}_{{\mathcal{S}_1}}^{\rm R}
  \qquad
  \text{for ${\mathcal{S}_1}\cap{\mathcal{S}_2}=\emptyset$. (See Fig.~\ref{fig:rr_disj}.)}
\end{equation}

\begin{figure}
  \caption{%
Successive operations of {\it I}-subtraction for subdiagram $\mathcal{S}_1$ 
and {\it R}-subtraction for subdiagram $\mathcal{S}_2$ when 
they satisfy $\mathcal{S}_1 \supset \mathcal{S}_2$. }
  \begin{center}
    \includegraphics[scale=0.8]{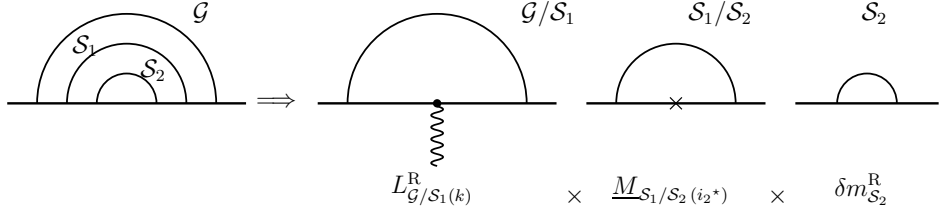}
  \end{center}
  \label{fig:ri_nest}
\end{figure}

\begin{figure}
  \caption{%
Successive operations of {\it I}-subtraction for subdiagram $\mathcal{S}_1$ 
and {\it R}-subtraction for subdiagram $\mathcal{S}_2$ when 
they are disjoint, $\mathcal{S}_1 \cap \mathcal{S}_2 = \emptyset$. }
  \begin{center}
    \includegraphics[scale=0.8]{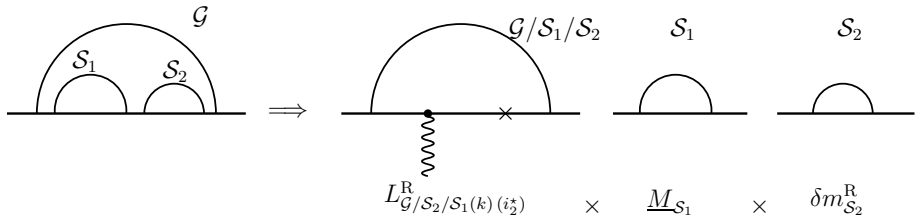}
  \end{center}
  \label{fig:ri_disj}
\end{figure}

Next we consider the case in which there are two operators 
$\mathbb{I}_{{\mathcal{S}_1}}$ and $\mathbb{R}_{{\mathcal{S}_2}}$. 
Since $\mathbb{I}_{{\mathcal{S}_1}}$ yields $\underline{M}_{{\mathcal{S}_1}}$ and
$\mathbb{R}_{{\mathcal{S}_2}}$ yields 
$\underline{M}_{\mathcal{G}/{\mathcal{S}_2}\,(i_2^{\star})}$, 
${\mathcal{S}_1}$ and $\mathcal{G}/{\mathcal{S}_2}$ 
must share some parts, 
{\it i.e.} 
${\mathcal{S}_1}\supset{\mathcal{S}_2}$ or 
${\mathcal{S}_1}\cap{\mathcal{S}_2}=\emptyset$. 
For the case ${\mathcal{S}_1}\supset{\mathcal{S}_2}$, 
the successive operations yield 
\begin{equation}
  \mathbb{R}_{{\mathcal{S}_2}}
  \mathbb{I}_{{\mathcal{S}_1}}
  \underline{M}_{\mathcal{G}}
  =
  \mathbb{R}_{{\mathcal{S}_2}}
  \left(
  {L}_{\mathcal{G}/{\mathcal{S}_1}}^{\rm R}\,
  \underline{M}_{{\mathcal{S}_1}}
  \right)
  =
  {L}_{\mathcal{G}/{\mathcal{S}_1}}^{\rm R}\,
  \underline{M}_{{\mathcal{S}_1}/{\mathcal{S}_2}\,(i_2^{\star})}\,
  {\delta m}_{{\mathcal{S}_2}}^{\rm R}
  \,,
  \qquad
  \text{(See Fig.~\ref{fig:ri_nest}.)}
\end{equation}
For the disjoint case ${\mathcal{S}_1}\cap{\mathcal{S}_2}=\emptyset$, 
it is given as 
\begin{equation}
  \mathbb{I}_{{\mathcal{S}_1}}
  \mathbb{R}_{{\mathcal{S}_2}}
  \underline{M}_{\mathcal{G}}
  =
  \mathbb{I}_{{\mathcal{S}_1}}
  \left(
  \underline{M}_{\mathcal{G}/{\mathcal{S}_2}\,(i_2^{\star})}\,
  {\delta m}_{{\mathcal{S}_2}}^{\rm R}
  \right)
  =
  {L}_{\mathcal{G}/{\mathcal{S}_2}/{\mathcal{S}_1}(k)\,(i_2^{\star})}^{\rm R}\,
  \underline{M}_{{\mathcal{S}_1}}\,
  {\delta m}_{{\mathcal{S}_2}}^{\rm R}
  \,,
  \quad
  \text{(See Fig.~\ref{fig:ri_disj}.)}  
\end{equation}
The case with ${\mathcal{S}_1}\subset{\mathcal{S}_2}$ is prohibited. 

To summarize, 
an IR-finite amplitude for a self-energy-like diagram 
$\mathcal{G}$ is obtained by subtracting away all IR divergences 
associated with its self-energy subdiagrams. 
They can be identified 
by applying two types of subtraction operations to the amplitude, 
called 
{\it I}-subtraction operation ($\mathbb{I}_{\mathcal{S}}$) 
and {\it R}-subtraction operation ($\mathbb{R}_{\mathcal{S}}$), 
for a set of annotated forests similar to Zimmermann's forests 
for UV divergences. 
An annotated forest consists of self-energy subdiagrams, to each 
of which a distinction of {\it I}-subtraction or {\it R}-subtraction 
operation is assigned. 
By construction, the annotation has to respect the assignment rules: 
\begin{itemize}
\item subdiagrams assigned to {\it I}-subtraction operations 
are included one another, 
\item a subdiagram $\mathcal{S}^\prime$ assigned to 
{\it R}-subtraction operation does not include a subdiagram 
$\mathcal{S}$ assigned to {\it I}-subtraction operation. 
\end{itemize}
The IR-finite amplitude is thus given by 
\begin{equation}
  \Delta M_{\mathcal{G}}
  =
  \underline{M}_{\mathcal{G}}
  +
  \sum_{{\ff}\in\tilde{\mathfrak{F}}}\,
  \left(-\mathbb{I}_{\mathcal{S}_i}\right)\cdots
  \left(-\mathbb{R}_{\mathcal{S}_j}\right)\cdots
  \underline{M}_{\mathcal{G}}
  \,,
\label{uvfinite}
\end{equation}
in which the summation is taken over all annotated forest 
${\ff} = \{ \mathcal{S}_i, \dots, \mathcal{S}_j, \dots \}$. 
(Recall that $\underline{M}_{\mathcal{G}}$ is UV-finite.)
The allowed combinations of the operators, 
including their ordering, are listed as follows:  
\begin{enumerate}[i) ]
\item 
$\mathbb{I}_{\mathcal{S}_i}\cdot\mathbb{I}_{\mathcal{S}_j}$ 
\hspace{2.2em} for $\mathcal{S}_i \subset \mathcal{S}_j$, 
\item 
$\mathbb{R}_{\mathcal{S}_i}\cdot\mathbb{R}_{\mathcal{S}_j}$ 
\hspace{1.5em} for $\mathcal{S}_i \supset \mathcal{S}_j$, 
\item 
$\mathbb{R}_{\mathcal{S}_i}\cdot\mathbb{R}_{\mathcal{S}_j} 
= \mathbb{R}_{\mathcal{S}_j}\cdot\mathbb{R}_{\mathcal{S}_i}$ 
\hspace{2em} for $\mathcal{S}_i \cap \mathcal{S}_j = \emptyset$, 
\item 
$\mathbb{R}_{\mathcal{S}_i}\cdot\mathbb{I}_{\mathcal{S}_j}$ 
\hspace{2em} for $\mathcal{S}_i \subset \mathcal{S}_j$, 
\item 
$\mathbb{I}_{\mathcal{S}_i}\cdot\mathbb{R}_{\mathcal{S}_j}$ 
\hspace{2em} for $\mathcal{S}_i \cap \mathcal{S}_j = \emptyset$. 
\end{enumerate}

It is noted that the operators 
$\mathbb{I}_{\mathcal{S}}$ and $\mathbb{R}_{\mathcal{S}}$ 
are regarded to act on the UV-finite amplitude 
$\underline{M}_{\mathcal{G}}$ 
in which UV divergences are subtracted away by {\it K}-operations. 
The {\it I}-/{\it R}-subtractions are introduced in such a way 
that they produce IR subtraction terms 
as products of UV-finite quantities. 
To be consistent with the formal prescription 
for successive operations, 
we have adopted the definitions of 
Eqs.~(\ref{eq:def:r-op}) and~(\ref{eq:def:i-op}). 
However, the identification of the IR subtraction terms can be 
carried out diagrammatically by the form of the diagram alone, 
and we do not have to examine the internal UV structure 
of the diagram nor the explicit expressions of integrands. 
Thus we introduce the regulations as a formal treatment 
that the {\it I}-/{\it R}-operators satisfy 
$\mathbb{I}_{\mathcal{S}} \mathbb{K}_{\mathcal{S}^\prime} = 0$ 
and
$\mathbb{R}_{\mathcal{S}} \mathbb{K}_{\mathcal{S}^\prime} = 0$
for {\it I}-({\it R}-)subtraction operator 
$\mathbb{I}_{\mathcal{S}}$ ($\mathbb{R}_{\mathcal{S}}$) 
for a subdiagram $\mathcal{S}$ 
and a {\it K}-operation $\mathbb{K}_{\mathcal{S}^\prime}$ 
for a subdiagram $\mathcal{S}^\prime$. 


\section{Construction of IR Subtraction Integrals}
\label{sec:construction}

In our numerical approach, the subtraction terms are given in 
the form of the parametric integrals that are defined over 
the same Feynman parameter space 
as that of the original diagram. 
These terms are prepared so that the IR singularities of 
the integrand are canceled at the same point. 
To attain this point-by-point subtraction of the singularity 
associated with the IR limit, 
we construct the IR subtraction term 
$L_{\mathcal{R}(k)}^{\rm R} \underline{M}_\mathcal{S}$ 
in such a way that the component terms 
$L_{\mathcal{R}(k)}^{\rm R}$ and $\underline{M}_\mathcal{S}$ 
expressed by separate parametric integrals are merged into 
a single integral by the Feynman integration formula 
\cite{Kinoshita:1990,Aoyama:2005kf}. 

In the previous section we have seen that the IR subtraction 
terms can be identified by annotated forests of the diagram 
$\mathcal{G}$. 
By the successive applications of {\it I}-/{\it R}-subtraction operations 
along the forest, 
the form of each term is obtained as a product of component terms, 
which are 
the residual part of the vertex renormalization constants 
and/or 
that of the mass renormalization constants, 
and 
the lower-order magnetic moment part $M$. 

In the actual construction, we follow the two-step procedure. 
First, we construct an integral that corresponds to a product 
of $M$, $\widetilde{L}\equiv L - L^{\rm UV}$ and/or 
$\widetilde{\delta m}={\delta m} - {\delta m}^{\rm UV}$. 
The location of component terms in the integration space, 
{\it i.e.} how the reduced diagrams of the components are 
embedded in the original diagram, is crucial to realize the 
point-wise subtraction. 
Here we introduce a tree representation of the annotated forest, 
called as an {\it annotated tree}, which provides a useful tool for 
identification of Feynman parameter assignment of the component 
integral. 

In the next step 
we subtract away the UV subdivergences of those terms 
by applying {\it K}-operations that are restricted onto the 
respective reduced diagrams. 
For this purpose, we introduce an embedding tree of the 
annotated tree, which corresponds to a particular combination 
of UV subdivergences of component terms. 


\subsection{Subtractive Integral in Feynman Parameter Space}
\label{sec:construction:integ}

The IR subtraction term is given as a product of component terms. 
Each component term is expressed as a parametric integral over 
the reduced diagram. 
Consider a case in which the IR subtraction term consists of 
two component terms of reduced diagrams $\mathcal{S}$ and 
$\mathcal{R}\equiv\mathcal{G}/\mathcal{S}$. 
Let $x_1, x_2,..., x_{n_\mathcal{S}}$ 
and $y_1, y_2,..., y_{n_\mathcal{R}}$ 
be the Feynman parameters of the diagram $\mathcal{S}$ and $\mathcal{R}$, 
and let $x_\mathcal{S} =\sum_i x_i$ and $y_\mathcal{R} =\sum y_i$, 
respectively. 
Then the IR subtraction term takes the form 
\begin{equation}
  \int (dx)_{\mathcal{S}}\,\frac{g[\mathcal{S}]}{V_{\mathcal{S}}^\alpha}
  \times
  \int (dy)_{\mathcal{R}}\,\frac{g[\mathcal{R}]}{V_{\mathcal{R}}^\beta}
  \,,
\label{irsub}
\end{equation}
where we denote the numerators of the part of integrands with 
the particular powers $\alpha$ and $\beta$ of $V$-function 
as 
$g[\mathcal{S}]$ for the component of the subdiagram $\mathcal{S}$, 
and 
$g[\mathcal{R}]$ for the component of the reduced diagram $\mathcal{R}$, 
respectively. 

By inserting the identity 
\begin{equation}
  1 = 
  \int_0^1\,\frac{ds}{s}\,\delta\left(1-\frac{x_{\mathcal{S}}}{s}\right)
  \,
  \int_0^1\,\frac{dt}{t}\,\delta\left(1-\frac{y_{\mathcal{R}}}{t}\right)
\end{equation}
in Eq.~(\ref{irsub}) and making use of the Feynman integral formula 
\begin{equation}
  \frac{\Gamma(k)}{A^k}\frac{\Gamma(l)}{B^l}
  \,
  =
  \Gamma(k+l) \int_0^1 ds\,dt\,\delta(1-s-t)\,
  \frac{s^{k-1} t^{l-1}}{(sA + tB)^{k+l}}
\end{equation}
the product of integrals in Eq.~(\ref{irsub}) is turned into 
a single integral over $\mathcal{G}$ of the form 
\begin{equation}
  \int (dz)_{\mathcal{G}}\,
  \frac{g[\mathcal{S}]\,g[\mathcal{R}]}
  {(V_\mathcal{S}+V_\mathcal{R})^{\alpha+\beta}}
  \,,
\end{equation}
where
\begin{align}
  z_i &= x_\mathcal{S} x_i \qquad \text{for $i=1,\dots,n_\mathcal{S}$,}
\nonumber \\
  z_i &= y_\mathcal{R} y_i \qquad \text{for $i=n_\mathcal{S} + 1,\dots,n_\mathcal{G}$.}
\end{align}
By this assignment the singularities in the Feynman parameter space 
of both the original integrand and that of the subtraction term 
cancel each other. 
This mechanism of cancellation is crucial for our subtraction procedure. 


\subsection{Tree Representation of Annotated Forest}
\label{sec:construction:tree}

\begin{figure}
\caption{An example of nested subdiagrams of a forest and 
the tree representation.}
\begin{center}
\subfigure[][]{%
  \includegraphics[scale=1.0]{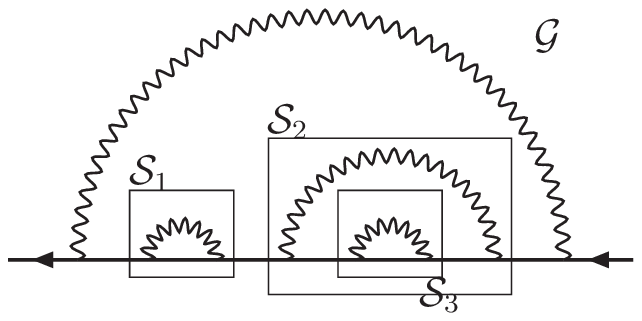}
}
\hskip1em
\subfigure[][]{%
  \includegraphics[scale=1.0]{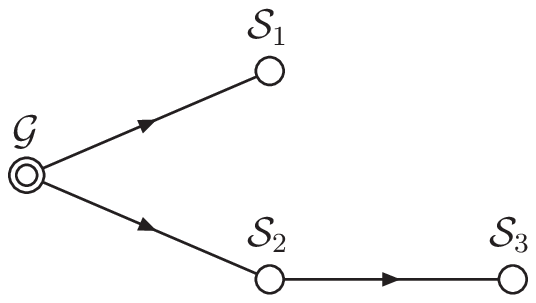}
}
\end{center}
\label{fig:tree}
\end{figure}

A forest is represented as a tree form that expresses the inclusion 
relation of the subdiagrams in the forest. 
(See Fig. \ref{fig:tree}.)
We assign each subdiagram to a node. 
If a subdiagram is included in another subdiagram $\mathcal{S}$, 
it is expressed as a child node of the node assigned 
to $\mathcal{S}$. 
We consider the diagram $\mathcal{G}$ itself as the root node 
of the tree. 
For later convenience, we denote the subdiagrams that contain 
another subdiagram $\mathcal{S}$ as (direct) ancestors of 
$\mathcal{S}$, and the subdiagrams that are included 
in $\mathcal{S}$ as descendants of $\mathcal{S}$. 

For the annotated forest, 
a distinct operation of {\it I}-subtraction or {\it R}-subtraction 
is assigned to the node. 
We call such a tree as {\it annotated tree} hereafter. 
Each node is then translated into the component term of the reduced 
diagram. 

For a simplest example in which $\mathcal{G}$ 
has a single subdiagram $\mathcal{S}$, 
the tree form of the forest is shown as follows: 
%
%
\begin{center}
        \includegraphics[scale=1.0]{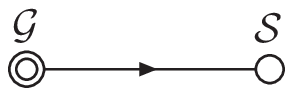}
\end{center}
When we consider the $\mathbb{I}_\mathcal{S}$ operation, 
the result is given as a product of a vertex renormalization constant 
${L}_{\mathcal{R}(k)}^{\rm R}$ for the part of the diagram 
$\mathcal{R}(k) \equiv \mathcal{G}/\mathcal{S}(k)$ 
and a magnetic moment part $\underline{M}_\mathcal{S}$ 
for the subdiagram $\mathcal{S}$. 
Those parts of subdiagrams are related to the nodes in the above tree 
labelled by $\mathcal{G}$ and $\mathcal{S}$, respectively. 
Thus we represent the assignment of components of the subtraction term 
graphically as below. 
%
%
\begin{center}
	\includegraphics[scale=1.0]{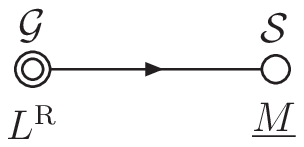}
\end{center}
When the {\it R}-subtraction $\mathbb{R}_\mathcal{S}$ 
is considered, the assignment is represented in a similar manner 
as follows. 
%
%
\begin{center}
	\includegraphics[scale=1.0]{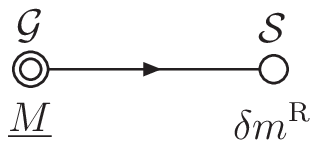}
\end{center}
Here, the subdiagram $\mathcal{S}$ referred by the right node 
is related to the residual self-mass term 
${\delta m}_\mathcal{S}^{\rm R}$, 
and the reduced diagram $\mathcal{G}/\mathcal{S}\,(i^\star)$ 
that corresponds to the left node 
is assigned to the magnetic moment part 
$\underline{M}_{\mathcal{G}/\mathcal{S}\,(i^\star)}$. 

This representation can be extended to more general cases. 
For an annotated forest that corresponds to a nested singularity, 
the successive operations of {\it I}-/{\it R}-subtractions in an 
appropriate order as described in Sec.~\ref{sec:ir:nested} 
is interpreted to extend the tree by following the respective 
process shown above. 
Then we obtain a tree representation of the annotated forest 
in which the nodes are assigned distinct types of 
$M$-, $L^{\rm R}$-, or ${\delta m}^{\rm R}$-nodes. 

The nodes of the tree are related to the reduced diagrams, and so 
the annotated tree has a direct interpretation 
to the IR subtraction term 
in the form of a product of the component terms for their 
respective reduced diagrams. 
Since the tree expresses the nested structure of subdiagrams in 
the forest, the relevant set of Feynman parameters for the 
component term, {\it i.e.} how the reduced subdiagram for the 
component term is embedded in the original diagram, can be 
easily read off from the tree. 
This feature is crucial in constructing the IR subtraction integral 
so that the point-wise subtraction of IR singularities is achieved. 

Thus far, the annotated tree provides a graphical representation of 
the annotated forest, and it has a one-to-one correspondence. 
It is readily translated into a symbolic form of the associated 
IR subtraction term, which is also significant for the residual 
renormalization step. 
The set of IR subtraction terms can be obtained by finding 
the set of annotated trees that have proper assignment of types 
to the nodes of tree consisted of self-energy subdiagrams. 

The rules for the annotated forests given in 
Sec.~\ref{sec:ir:nested} 
are reflected to those for the assignment of types of components 
in the annotated trees. 
They are summarized as follows: 
\begin{enumerate}
\item There is one and only one node to which the magnetic moment part 
$\underline{M}$ is assigned. 
\item The nodes that are assigned to the {\it I}-subtraction, 
${L}^{\rm R}$, 
are restricted to the ancestor nodes of the magnetic moment part. 
\item The nodes that are assigned to the residual self-mass 
subtraction, 
${\delta m}^{\rm R}$, 
do not appear as the ancestor nodes of the magnetic moment part. 
\end{enumerate}
It turns out that to fulfill these rules the assignment is 
uniquely determined once a node is chosen for the magnetic moment part. 
We first pick up a node that is assigned to the magnetic part 
$\underline{M}$, 
and then the nodes that lie as ancestors of the $M$-node 
are associated with the {\it I}-subtractions. 
The remaining nodes are assigned to the {\it R}-subtractions. 


\subsection{Subtraction of UV Subdivergences}
\label{sec:construction:uvsubd}

We now proceed to the treatment of UV subdivergences 
in the components of IR subtraction terms. 
For example, the residual part of vertex renormalization constant 
$L^{\rm R}$ is given in the form 
\begin{equation}
  L_{\mathcal{R}(k)}^{\rm R} 
  = 
  L_{\mathcal{R}(k)} - L_{\mathcal{R}(k)}^{\rm UV} 
  + 
  \sum_{f\in\mathfrak{F}}\,
  \prod_{\mathcal{S}^\prime\in f}
  \left(-\mathbb{K}_{\mathcal{S}^\prime}\right)\,
  \widetilde{L}_{\mathcal{R}(k)} 
  \,.
\end{equation}
We have introduced a term 
$\widetilde{L}_{\mathcal{R}(k)} \equiv L_{\mathcal{R}(k)} - L_{\mathcal{R}(k)}^{\rm UV}$ 
in the construction of the subtraction term which is given from 
$L_{\mathcal{R}(k)}$ 
by dropping the most-contracted term 
$L_{\mathcal{R}(k)}^{\rm UV}$. 
The divergences that originate from the substructure 
have to be subtracted away that are dealt with {\it K}-operations 
for the forests $f$ of $\mathcal{R}(k)$. 
In this section we show that these subdivergences can be handled 
in a systematic way and present it as a general scheme. 

The IR subtraction term constructed in the previous section is 
originally given as a product of separate integrals defined 
on the respective component diagrams. 
As an example, the term 
${L}_{\mathcal{R}(k)}^{\rm R} \underline{M}_\mathcal{S}$ is made of 
the vertex renormalization constant defined for the vertex diagram 
$\mathcal{R}(k)$ and the magnetic moment part defined for the diagram 
$\mathcal{S}$. 
Those components have their own divergent subdiagrams. 

It is shown that these subdiagrams of separate components can 
be mapped to subdiagrams of the original diagram $\mathcal{G}$. 
A subdiagram $\tilde{\mathcal{S}}$ of a component diagram 
is itself a subdiagram of $\mathcal{G}$, or 
$\tilde{\mathcal{S}}$ is obtained as a reduced diagram by 
shrinking other component diagrams. 

Thus, it can be recognized graphically in a way that 
a forest of a component diagram is related to a forest 
of the original diagram. 
The tree form of the forest of the component diagram can 
be embedded as a subtree of the original annotated forest 
that corresponds to the IR subtraction term 
by mapping the elements of the subtree as that of the 
original forest. 
We call such a tree that corresponds to UV subdivergence 
of certain components the embedding tree. 
\begin{figure}
  \caption{A UV divergent subdiagram $\tilde{\mathcal{S}}$ in 
    the annotated forest $\{\mathcal{G},\mathcal{S}\}$.}
  \begin{center}
    \includegraphics[scale=1.0]{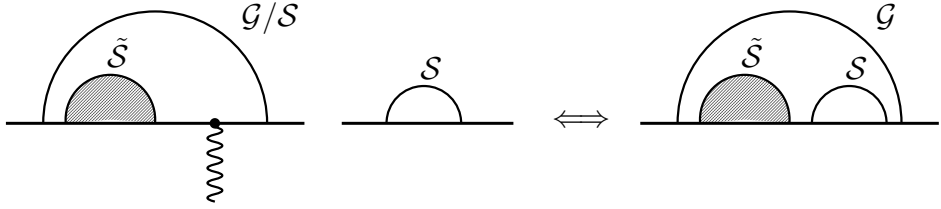}
  \end{center}
  \label{fig:ir_subd}
\end{figure}
For example, consider a case in which the component diagram 
$\mathcal{R} = \mathcal{G}/\mathcal{S}$ has a forest 
consisting of a subdiagram $\tilde{\mathcal{S}}$ as shown in 
Fig.~\ref{fig:ir_subd}. The tree form of this forest 
%
%
\begin{center}
	\includegraphics[scale=1.0]{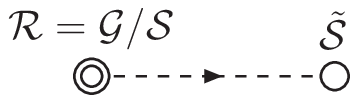}
\end{center}
is embedded in the annotated forest
%
%
\begin{center}
	\includegraphics[scale=1.0]{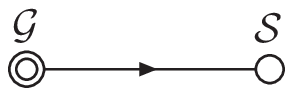}
\end{center}
which reduces to the forest of the whole diagram $\mathcal{G}$ 
shown as follows.
%
%
\begin{center}
	\includegraphics[scale=1.0]{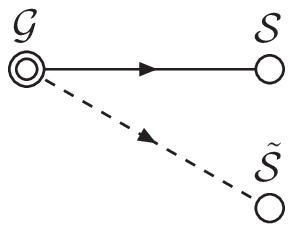}
\end{center}
It is noted that the component diagrams are disjoint 
by construction and therefore the subdiagrams of 
different component diagrams are not overlapping. 

The UV subdivergences contained in the product of the components 
are identified by forests of the original diagram $\mathcal{G}$. 
For the annotated forest ${\ff}$ that corresponds to the 
IR subtraction term, consider a forest $f^\prime$ that includes 
${\ff}$, in the sense that all subdiagrams of ${\ff}$ 
are also the elements of $f^\prime$. 
Then the subdiagrams in $f^\prime - {\ff}$, 
{\it i.e.} the subdiagrams of $f^\prime$ that are not in ${\ff}$, 
are related to the subdivergences of the component diagram 
to which they belong respectively. 
The subtraction term of the UV subdivergence is constructed 
by applying {\it K}-operations for these subdiagrams restricted onto 
the corresponding component diagram. 
For example, $\mathbb{K}_{\tilde{\mathcal{S}}}$ acts 
on the component $\mathcal{G}/\mathcal{S}$ in Fig.~(\ref{fig:ir_subd}). 

Therefore, the UV subdivergences for an IR subtraction term 
that originate from the distinct component terms are 
synthetically identified by the forests of the original 
diagram $\mathcal{G}$. 
However, 
not all the subdiagrams of $\mathcal{G}$ cause divergences 
of the IR subtraction term related to the annotated forest 
${\ff}$. 
Consider a case in which a vertex subdiagram $\mathcal{S}^\prime$ 
includes a self-energy subdiagram $\mathcal{S}$, 
{\it i.e.} $\mathcal{S}^\prime\supset\mathcal{S}$. 
For the {\it I}-subtraction associated with $\mathcal{S}$, 
the subdiagram $\mathcal{S}$ is replaced by an external vertex. 
Thus $\mathcal{S}^\prime$ acquires the extra vertex 
in the reduced diagram $\mathcal{G}/\mathcal{S}(k)$ 
and becomes irrelevant for the UV divergence. 
We have to exclude such cases from the identification of 
subdivergences. 

Another rule must be imposed in relation to the residual 
self-mass subtraction. 
First consider a nested UV divergences in which self-energy 
subdiagram $\mathcal{S}$ is included 
in a vertex or another self-energy subdiagram $\tilde{\mathcal{S}}$. 
There would be a UV subtraction term denoted as 
${\delta m}_{\mathcal{S}}^{\rm UV} L_{\tilde{\mathcal{S}}/\mathcal{S}\,(i^\star)}^{\rm UV}$ 
for the vertex subdiagram, 
or 
${\delta m}_{\mathcal{S}}^{\rm UV} B_{\tilde{\mathcal{S}}/\mathcal{S}\,(i^\star)}^{\rm UV}$ 
for the self-energy subdiagram 
in the prescription of {\it K}-operation. 
However, by construction of {\it K}-operation it is found that 
those terms that involve the most-contracted part of 
the vertex renormalization constants 
or wave-function renormalization constants 
with mass insertions do not actually appear. 
Accordingly, we also have to omit the residual self-mass term 
${\delta m}_{\mathcal{S}}^{\rm R} L_{\tilde{\mathcal{S}}/\mathcal{S}\,(i^\star)}^{\rm UV}$ 
or 
${\delta m}_{\mathcal{S}}^{\rm R} B_{\tilde{\mathcal{S}}/\mathcal{S}\,(i^\star)}^{\rm UV}$, 
respectively. 
They would be realized as UV subdivergences of the component 
$\tilde{\mathcal{S}}/\mathcal{S}$. 


\section{Procedure and Implementation}
\label{sec:proc}

In this section we describe the flow of process to generate 
the numerical integration code for IR subtraction terms of a 
{\it q-type} diagram from its representation. 

We first identify the IR divergent parts based on the forest 
structure of the diagram and provide the set of subtraction terms 
as annotated forests. 
Next we identify the UV subdivergences of these IR subtraction terms. 
The subtraction terms for these subdivergences are constructed 
by applying {\it K}-operations to the IR subtraction integrals. 

\begin{figure}
  \rotatebox{90}{\includegraphics[scale=0.65]{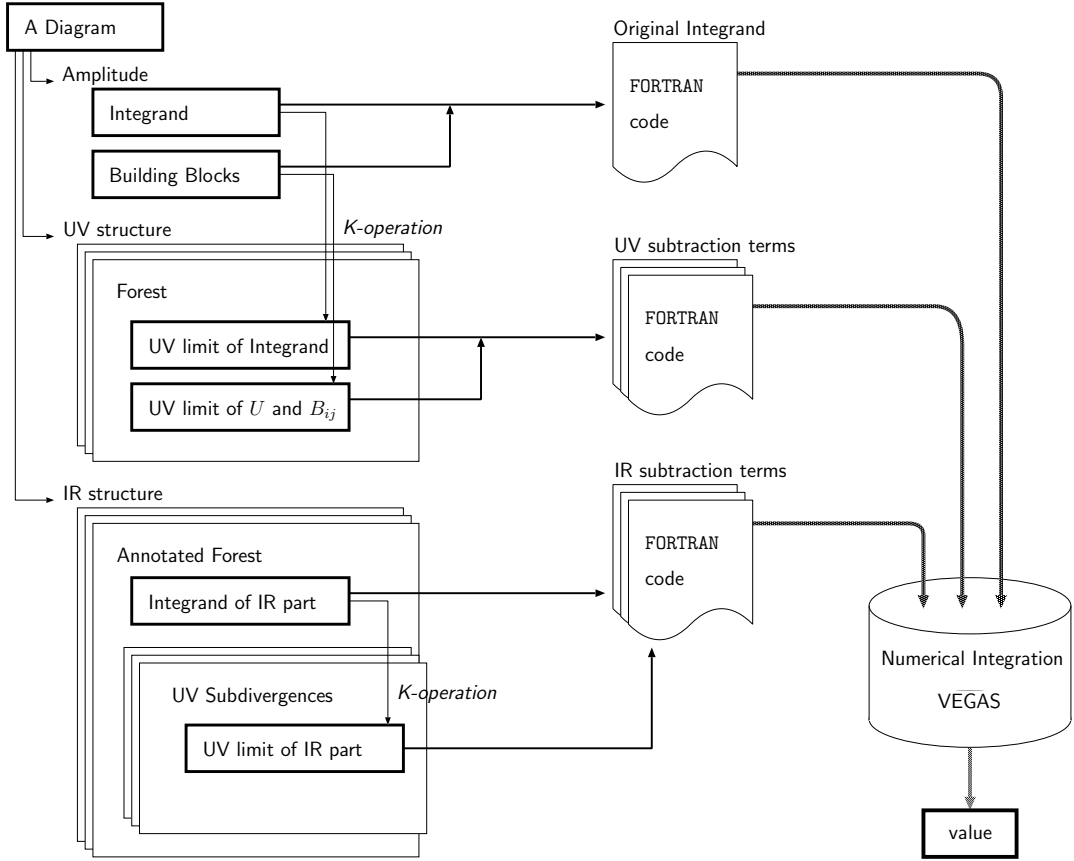}}
\caption{Flow of automated code generation.}
\label{fig:flow}
\end{figure}

The subtraction terms of IR divergences and their UV subdivergences 
thus constructed are turned into FORTRAN codes. 
They are merged to the numerical calculation codes 
for UV finite amplitude generated by the scheme developed in 
Ref.~\cite{Aoyama:2005kf}. 
The entire flow of process is depicted in Fig.~\ref{fig:flow}. 
The numerical integration codes are readily integrated 
by VEGAS to produce a finite amplitude for the diagram. 
Some of the steps are common to the procedure for UV subtraction 
described in Ref.~\cite{Aoyama:2005kf}. 

\subsection{Finding Forest Structures}
\label{sec:proc:find-forest}

The types of subdiagrams that are relevant to the IR subtraction 
term and their UV subdivergences are self-energy subdiagrams and 
vertex subdiagrams. 
For a {\it q-type} diagram a subdiagram of these types is 
represented by a segment of the lepton path. 
All subdiagrams of a diagram are thus found by examining all 
segments that correspond to the connected one-particle 
irreducible subdiagrams of vertex type or self-energy type. 

The inclusion relation of the subdiagrams are thus mapped to 
that of the segments. 
Once it is found, 
the whole set of (normal) forests is found by generating all 
possible combinations of subdiagrams and disregarding any 
combinations in which a pair of elements are overlapping. 

For each forest we examine the nest structure of subdiagrams 
and describe it in a tree form whose root node is associated 
with the whole diagram $\mathcal{G}$. 

\subsection{Finding IR Subtraction Term}
\label{sec:proc:find-ir}

The IR divergences of a {\it q-type} diagram emerge in relation to 
subdiagrams of the self-energy type. 
To identify the divergence structure, we first pick out the 
forests that consist of only self-energy subdiagrams. 
Suppose they are expressed in a tree form. 
Next, we assign to each node of the tree the type of component 
of subtraction term among the choices: 
the residual part of vertex renormalization constant ($L^{\rm R}$) 
that is related to {\it I}-subtraction, 
the residual self-mass part (${\delta m}^{\rm R}$) 
that is related to {\it R}-subtraction, 
and the magnetic moment part ($\underline{M}$). 
The assignment that fulfills the rules in Sec.~\ref{sec:ir} 
can be found by following steps. 
\begin{enumerate}
\item Choose a node for the magnetic moment part 
(denoted by $M$-node). 
\item The ancestor nodes of $M$-node are assigned to the $L^{\rm R}$-part 
(denoted by $L^{\rm R}$-node). 
\item Other nodes are assigned to the residual self-mass parts 
(denoted by ${\delta m}^{\rm R}$-node). 
\end{enumerate}
We call the forest with proper assignment of types to the nodes 
of its tree representation as the annotated forest. 
Each annotated forest corresponds to a IR subtraction term. 

The expression of the IR subtraction term associated with the 
annotated forest is then given in a form of 
a product of component terms 
that are related to the nodes of the tree. 
The component term of a node $\mathcal{S}$ is found by following 
the rules below. 
\begin{itemize}
\item An $L^{\rm R}$-node corresponds to the residual part of 
the vertex renormalization constant $L^{\rm R}_{\mathcal{R}(k)}$ 
for the diagram $\mathcal{R}(k)$. 
It is obtained from $\mathcal{S}$ by replacing the inner subdiagram 
that is assigned to $M$-node or $L^{\rm R}$-node by a vertex $k$, 
and by shrinking the subdiagrams of child nodes to points. 
\item A ${\delta m}^{\rm R}$-node corresponds to the residual part 
of the mass renormalization constant ${\delta m}^{\rm R}_{\mathcal{R}}$ 
for the diagram $\mathcal{R}$. 
It is obtained from $\mathcal{S}$ by shrinking the subdiagrams 
of child nodes to points. 
\item An $M$-node corresponds to the magnetic moment part 
$\underline{M}_\mathcal{R}$ 
for the diagram $\mathcal{R}$ obtained from $\mathcal{S}$ 
by shrinking the subdiagrams of child nodes to points. 
\end{itemize}
As an example, consider an annotated tree shown as 
%
%
\begin{center}
	\includegraphics[scale=1.0]{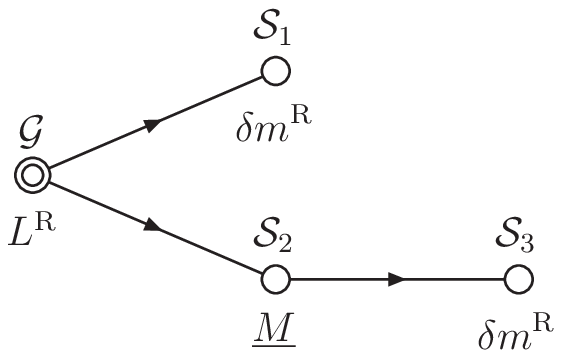}
\end{center}
It corresponds to the IR subtraction term 
\begin{equation}
	L^{\rm R}_{\mathcal{G}/\mathcal{S}_1/\mathcal{S}_2(k_2)\,({i_1}^\star)}\ 
	\delta m^{\rm R}_{\mathcal{S}_1}\ 
	M_{\mathcal{S}_2/\mathcal{S}_3\,({i_3}^\star)}\ 
	\delta m^{\rm R}_{\mathcal{S}_3}.
\end{equation}

Since the mass renormalization constant of second order 
$\delta m_2$ satisfies 
$\delta m_2^{\rm UV} = \delta m_2$, 
the residual self-mass subtraction for the second-order 
self-energy subdiagram is not necessary. 
We omit the annotated forests that involve such cases. 

An example of a complete set of IR subtraction terms 
for an eighth-order diagram 
identified in the above procedure is shown 
in Appendix~\ref{sec:example}.

\subsection{Constructing IR Subtraction Integral}
\label{sec:proc:ir-integral}

The integrand of the IR subtraction term associated with 
an annotated forest ${\ff}$ is constructed by the following 
steps. 
First we identify the position of each component term in
the original diagram $\mathcal{G}$, 
and find the respective sets 
of Feynman parameters for those components. 
This step is crucial for our numerical method based on the 
point-by-point subtraction. 
Then we follow these steps: 
\begin{enumerate}
\item Construct the integrand as a product of those of component terms. 
\item Find the building blocks, $B_{ij}$, $U$, $C_{ij}$ and $A_i$. 
\item Construct the $V$-function in a form that guarantees the 
analytical factorization of the subtraction integral into components. 
\end{enumerate} 

\subsubsection{Integrand}
\label{sec:proc:ir-integral:integrand}

The numerator of the integrand of the subtraction term is given 
as a product of those of component terms. 
We first consider the full part of the component integrands, 
the vertex renormalization constant $L$, the mass renormalization 
constant $\delta m$, and the magnetic part $M$, 
before dropping other than the residual part. 

For example, consider an annotated forest consisted of 
an {\it I}-subtraction for the subdiagram $\mathcal{S}$. 
We construct the subtraction term of the form of product 
\begin{equation}
	\widetilde{L}_{\mathcal{R}(k)}\ M_\mathcal{S}. 
\end{equation}
The numerator of the integrand 
(except for the numerical constants) 
is thus given by 
\begin{equation}
	\Tr_{\mathcal{R}(k)} \Biggl\{
	P_\nu \gamma^\alpha (\sla{p}-m) \cdots \sla{p}_k^\nu 
	\cdot
	\Tr_{\mathcal{S}} \biggl\{
	P \gamma^\beta (\sla{p}-m) \cdots \gamma_\beta 
	\biggr\}
	\cdot
	(\sla{p}-m) \cdots \gamma_{\alpha^\prime}
	\Biggr\}, 
\label{eq:integrand-example}
\end{equation}
where the labels $\mathcal{R}(k)$ and $\mathcal{S}$ of the traces 
indicate that the traces are taken separately for the components 
in $\mathcal{R}(k)$ and $\mathcal{S}$. 
$P_\nu$ and $P$ are the projection operators for 
the vertex renormalization constant and 
the magnetic moment, respectively, 
and the term $\sla{p}_k$ refers to the external vertex $k$. 

The expression (\ref{eq:integrand-example}) is analogous to 
the integrand of the unrenormalized amplitude of the original 
diagram $\mathcal{G}$ except for these points: 
\begin{itemize}
\item There are multiple traces, each of which corresponds to 
the respective component term. 
\item Each part has appropriate projection operator inserted. 
\item For the $L^{\rm R}$-parts associated 
with the {\it I}-subtractions, extra vertices are inserted. 
\end{itemize}
On the other hand, the pattern of contractions of the internal 
vertices by the photon propagators is the same as that of 
the original diagram $\mathcal{G}$ given by the pairings 
of the diagram representation. 
Thus we construct the integrand of the subtraction term 
based on that of the unrenormalized amplitude by incorporating 
modifications listed above. 

We perform the trace operations, and turn the integrand 
into a Feynman parametric form by carrying out the momentum 
integration analytically using the integration table. 
The result is expressed as a polynomial of the symbols called 
the building blocks, $B_{ij}$, $U$, $C_{ij}$, $A_i$ and $V$, 
each of which is given as a function of Feynman parameters. 

We have to remove the leading overall UV divergent parts of 
the component terms $L$ and $\delta m$ to leave their residual 
parts. 
It is achieved by dropping the most-contracted term of 
the component integral which can be identified by simple 
power counting. 

\subsubsection{Building Blocks}
\label{sec:proc:ir-integral:buildingblocks}

The integrand of the IR subtraction term is constructed as 
a product of component terms, each of which is defined on 
the distinct set of Feynman parameters and given in the 
form of rational function of building blocks that are also defined 
on the respective component diagram. 
However, we need not prepare these building blocks separately 
if we recall that the building blocks $B_{ij}$ and $U$ 
factorize exactly in the UV limit. 

In the UV limit associated with a subdiagram $\mathcal{S}$, 
the building blocks $B_{ij}$ and $U$ factorize in the form 
\begin{align}
	U_{\mathcal{G}} 
	& \to 
	U_{\mathcal{G}/\mathcal{S}} \cdot U_{\mathcal{S}}, \\
	\left. \frac{B_{ij}}{U} \right|_\mathcal{G} 
	& \to 
	\left. \frac{B_{ij}}{U} \right|_\mathcal{S} 
	\qquad\text{for\ }i,j \in \mathcal{S}, \\
	\left. \frac{B_{ij}}{U} \right|_\mathcal{G} 
	& \to 
	\left. \frac{B_{ij}}{U} \right|_{\mathcal{G}/\mathcal{S}}
	\qquad\text{for\ }i,j \in \mathcal{G}/\mathcal{S}, 
\end{align}
where the labels 
$\mathcal{G}$, $\mathcal{S}$, and $\mathcal{G}/\mathcal{S}$ 
denote that the quantities are defined on the respective diagrams. 

Therefore, we are able to use the building blocks $B_{ij}$ and $U$ 
obtained in the UV limit associated with the same forest $f$. 
We have to eliminate unwanted contributions by explicitly putting 
the elements of $B_{ij}$ to zero for 
$i\in\mathcal{S}$ and $j\not\in\mathcal{S}$ to ensure that 
the integrand originates from the separate integrals. 

\subsubsection{$V$-function}
\label{sec:proc:ir-integral:v-function}

In order that the exact factorization of the integrand holds, 
the $V$-function in the denominator of the integrand is 
constructed in the form 
\begin{equation}
  V \to \sum_{\mathcal{S}} V_\mathcal{S}, 
\end{equation}
where the sum is taken over the component diagrams. 

\subsection{Subtraction Terms for UV Subdivergences}
\label{sec:proc:uvsubd}

The integral forms of the subtraction terms for UV subdivergences 
that originate from the components of the IR subtraction term 
are constructed by the following steps. 
We assume that the IR subtraction term is associated with the 
annotated forest ${\ff}$. 

\subsubsection{Identification of Subdivergences}
\label{sec:proc:uvsubd:terms}

The set of UV subdivergences to the annotated forest ${\ff}$ 
is given by finding forests $f^\prime$ that include ${\ff}$. 
The emergence of UV subdivergences are associated with the 
subdiagrams in $(f^\prime - {\ff})$, {\it i.e.} the subdiagrams of 
$f^\prime$ that are not the elements of ${\ff}$. 
By the tree forms of the forests $f^\prime$ and ${\ff}$ 
we identify which component of IR subtraction term each of 
these subdiagrams belongs to. 
We check if these subdiagrams in the component diagrams 
actually cause the divergences, and disregard the cases 
described in Sec.~\ref{sec:construction:uvsubd}. 

\subsubsection{Integrand}
\label{sec:proc:uvsubd:integrand}

The integrand of the subtraction term for the UV subdivergence 
is obtained by applying the set of {\it K}-operations 
identified in the previous step. 
The individual {\it K}-operation is carried out by the power counting 
in a similar manner as the ordinal {\it K}-operations except that 
the domain of operation is restricted to the corresponding 
component diagram that the subdiagram belongs to. 
When there are more than one subdiagram, the {\it K}-operations are 
applied successively. The order of applications to different 
components is indifferent to the result. 
For a proper definition of the integrand Feynman cutoff must be
introduced for UV-divergent subdiagrams.
Of course, the cutoff dependence cancels out in the end.

\subsubsection{Building Blocks}
\label{sec:proc:uvsubd:buildingblocks}

For each component diagram, the building blocks for the 
subtraction term of the UV subdivergence are obtained 
in the factorized form for the respective UV limit. 
Recall that the building blocks of the IR subtraction term for 
the annotated forest ${\ff}$ is given as those obtained 
for the UV subtraction term associated with the same forest. 
Thus they reduce to the building blocks $B_{ij}$ and $U$ 
in the UV limit of the forest $f^\prime$. 

\subsection{Implementation}
\label{sec:proc:implementation}

We implemented the steps described above as separate Perl programs 
that use internally FORM \cite{Vermaseren:2000nd} and Maple. 
These symbolic manipulation programs 
take traces, 
project out the magnetic moment, 
perform analytic integration over momentum variables 
by means of home-made integration tables written in FORM, 
carry out inversion of matrices which creates $B_{ij}$ and $U$, 
and execute $K$-operations. 
The programs for the IR subtraction part are integrated with 
the programs that generate the codes for UV-finite amplitudes 
previously developed \cite{Aoyama:2005kf}, 
to form the automated code-generating system for the finite 
amplitude free from both UV and IR divergences. 

The code-generating system takes the name of the diagram 
and finds the corresponding 
single-line expression of the diagram from the table
prepared beforehand. 
Then it generates the numerical integration code in FORTRAN format 
that consists of the unrenormalized amplitude and appropriate 
subtraction terms. 
The FORTRAN code is readily integrated by a numerical integration program, 
to which purpose we use VEGAS \cite{vegas}, 
an adaptive Monte-Carlo integration routine. 

The whole steps are controlled by {\tt make} utility and a shell 
script. 
It is noted that some steps are independent with each other so that 
they can be performed simultaneously. 
By specifying appropriate options to {\tt make}, those operations 
are performed in parallel. 
This feature is beneficial in the 
present multi-processor/multi-core computer environment. 


\section{Conclusion and Discussion}
\label{sec:conclusion}

In this article we described a systematic procedure for the 
construction of IR subtraction terms in the calculation of 
lepton anomalous magnetic moment. 
Our present concern is the type of diagrams without internal 
lepton loops (called {\it q-type} diagrams) that account for 
a significant fraction and the most difficult portion of
calculation of QED corrections to lepton $g\!-\!2$. 
We presented a procedure for the automated generation of the 
numerical integration code, which, together with the previously 
developed automation scheme for UV-finite amplitude, provides 
prescription for the fully-renormalized and finite amplitude. 

We implemented the code-generating system based on our procedure 
as a set of Perl programs that work with the helps of 
symbolic manipulation systems, FORM and Maple. 
The entire flow of process is managed by {\tt make} utility 
and a shell script. 
It generates numerical integration code in FORTRAN format, which 
is readily integrated by numerical integration system, VEGAS. 
Our implementation is applicable to any order of diagrams of 
{\it q-type}. 

For the purpose of debugging of our
automation system 
we have applied it to the evaluation of known diagrams
of sixth- and eighth-orders. 
The sixth-order contribution of {\it q-type} diagrams 
comes from fifty vertex diagrams, which are reduced 
to eight self-energy-like diagrams 
taking account of
the Ward-Takahashi identity and the time-reversal symmetry.
It takes just one minute to create all eight FORTRAN programs 
for $M_{6\alpha}$ ($\alpha = a, b, \dots, h$) 
on {\sl hp}'s Alpha machine. 
Numerical evaluation was carried out on RIKEN's cluster system (RSCC). 
The result obtained 
is, after incorporating the residual renormalization, 
\begin{equation}
	A_1^{(6)}\text{(q-type)} = 0.9052~(11), 
\label{eq:6th-numerical}
\end{equation}
which reproduces 
the exact result \cite{Laporta:1996mq}
quite satisfactorily 
\begin{equation}
	A_1^{(6)}\text{(q-type)}_{\rm exact} = 0.904979\dots.
\end{equation}
The value (\ref{eq:6th-numerical}) was obtained by
computation by VEGAS of 2 -- 6 hours wall-clock time 
with 16 Xeon-CPU's for each diagram, which sampled 
one hundred million points per iteration and 450 iterations. 

The eighth-order {\it q-type} diagrams consist of 518 vertex diagrams, 
which are reduced to 47 self-energy-like diagrams 
with the help of the Ward-Takahashi identity and the time-reversal symmetry. 
The entire 47 program sets are generated 
by our automated code-generating system 
in less than ten minutes on {\sl hp}'s Alpha machine. 
The numerical evaluation is, however, quite non-trivial and requires 
huge computational resources. 
For the preliminary evaluation we have used 64 to 256 CPU's 
per diagram for a few months on RSCC 
to reach the relative uncertainty of about 3 \%. 

One unexpected byproduct is that it
revealed an inconsistency in 
the treatment of IR subtraction terms in the previous calculation of 
the eighth-order {\it q-type} diagrams \cite{Aoyama:2007dv}. 
With this inconsistency resolved, the new and old calculations 
agreed within the numerical precision employed. 
Note that this is the first time that the complete 
eighth-order contribution to $g\!-\!2$ has been calculated 
by more than one independent method. 
The agreement of two independent calculations puts the eighth-order 
contribution to $a_e$ on a firm ground, and advances the 
test of QED  and the determination of 
the fine structure constant derived from $a_e$ 
to a higher level of precision. 

These tests have confirmed the validity of our automation system, 
and encouraged us to tackle the evaluation of
the tenth-order contribution with confidence. 
Of 6354 {\it q-type} vertex diagrams 
that contribute to the tenth-order term, 
2232 diagrams that have only vertex subdiagrams have already been 
evaluated \cite{Aoyama:2005kf}.
For the remaining 4122 diagrams that have self-energy subdiagrams 
and thus suffer from the infrared singularity, we 
are evaluating their Ward-Takahashi version (254 diagrams) 
by the automation system described in this article.
It seems thus far that the numerical results of these diagrams 
are finite and our subtraction scheme works as expected. 
Some diagrams show somewhat unstable behavior during the evaluation 
which seems to indicate 
the presence of a severe digit-deficiency problem\footnote{%
For a detailed discussion of digit-deficiency problem, see Appendix B
of Ref.~\cite{Kinoshita:2004wi}.}.
To deal with such cases we
have modified the code so that we can carry out the integration 
with extended numerical precision. 

It is important to note that our
new subtraction scheme for IR divergences differs
from the scheme described in Ref.~\cite{Kinoshita:1990} 
by finite quantity. 
In the previous approach, the IR subtraction term was chosen 
for a self-energy subdiagram $\mathcal{S}$ in the form 
of Eq.~(\ref{eq:qedsub}) defined by the {\it I}-operation alone. 
In the new scheme, we have chosen 
the form ${L}_{\mathcal{R}(k)}^{\rm R}$ for the subtraction term 
defined as the residual part of $L_{\mathcal{R}(k)}$ 
other than the UV divergences. 
This choice of term includes a finite part $\Delta L$ 
in the previous calculations.  
Thus the subtraction of the nested IR singularity is 
carried out differently.
The latter part of Eq.~(\ref{eq:qedsub}) is related 
to our residual self-mass subtraction, but the form of term 
$M_{\mathcal{R}^\star}[I]$ was chosen through the careful 
inspection of the divergent structure of the particular diagram. 

Therefore the previous scheme and the new scheme yield different 
integrals for diagrams containing self-energy subdiagrams. 
The difference can be traced analytically.
In the eighth-order case it is confirmed by numerical calculation. 
The difference is, of course, to be compensated by the different 
set of terms for the residual renormalization step. 

In this series of articles we have focused on the particular 
type of diagrams that have no lepton loops ({\it q-type}). 
However, our procedure is readily applicable to other types of 
diagrams that are obtained by inserting vacuum polarizations 
to {\it q-type} diagrams. 
Such sets of diagrams of the tenth order are 
\cite{Kinoshita:2005sm,Aoyama:2005kf}
\begin{itemize}
\item Set III(a) obtained from the sixth-order {\it q-type} diagrams by 
inserting two second-order vacuum polarizations, 
\item Set III(b) obtained from the sixth-order {\it q-type} diagrams by 
inserting a fourth-order vacuum polarization, 
\item Set IV obtained from the eighth-order {\it q-type} diagrams by 
inserting a second-order vacuum polarization. 
\end{itemize}
The automated code generation for these sets is accomplished 
with only small modifications to the code. 
The integration codes have already been obtained, and 
the numerical evaluation has been carried out. 
The result will be reported elsewhere. 


\begin{acknowledgments}
This work is supported in part by 
JSPS Grant-in-Aid for Scientific Research (C)19540322. 
T. K.'s work is supported by the U. S. National Science Foundation
under Grant PHY-0355005.
M. H.'s work is supported in part by 
JSPS and the French Ministry of Foreign Affairs under the 
Japan-France Integrated Action Program (SAKURA). 
The numerical calculation has been performed on the RIKEN 
Super Combined Cluster System (RSCC).
\end{acknowledgments}

\appendix

\section{Identification of IR subtraction terms: an example}
\label{sec:example}

\begin{figure}
  \begin{center}
    \includegraphics[scale=1.0]{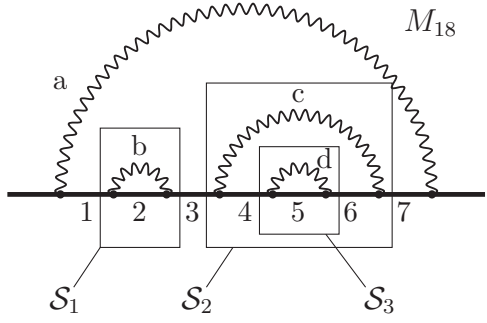}
  \end{center}
  \caption{%
An eighth-order diagram $M_{18}$. 
The numerals denote the indices of lepton lines 
and the roman alphabets denote the indices of photon lines.
The diagram $M_{18}$ has three self-energy subdiagrams denoted by 
$\mathcal{S}_1$, $\mathcal{S}_2$, and $\mathcal{S}_3$. 
}
\label{fig:m18}
\end{figure}

As a demonstration of the IR subtraction scheme, 
we present the identification of IR subtraction terms for 
an eighth-order diagram, $M_{18}$, shown in Fig.~\ref{fig:m18}. 

The diagram $M_{18}$ has three self-energy subdiagrams 
denoted by $\mathcal{S}_1$, $\mathcal{S}_2$, and $\mathcal{S}_3$. 
The number of combinations of those subdiagrams (forests) is seven, 
and the number of ways to assign the types 
($M$, $L^{\rm R}$, or ${\delta m}^{\rm R}$) to the nodes is 19. 
Dropping the cases that contain the second-order residual self-mass 
subtractions ${\delta m}_2^{\rm R} = 0$, 
it is found that the IR subtraction terms are given by 
the following six cases. 
On the left-hand side is shown the annotated tree, and 
the figure on the right-hand side expresses 
the reduced subdiagrams of the component terms. 

As a convention we denote the subtraction term by the symbol $I$ and/or $R$ 
with suffixes that refer to the indices of lepton lines. 
They are the indices to lepton lines contained in $L^{\rm R}$ for 
{\it I}-subtraction, and the range of indices to lepton lines 
indicated by the left-most and the right-most indices 
contained in ${\delta m}^{\rm R}$ for {\it R}-subtraction. 

\begin{itemize}
\item $I_{134567}$ \\
\includegraphics[scale=0.9]{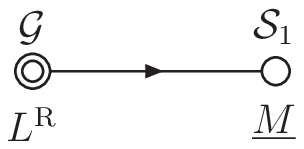} \hskip8em
\includegraphics[scale=0.6]{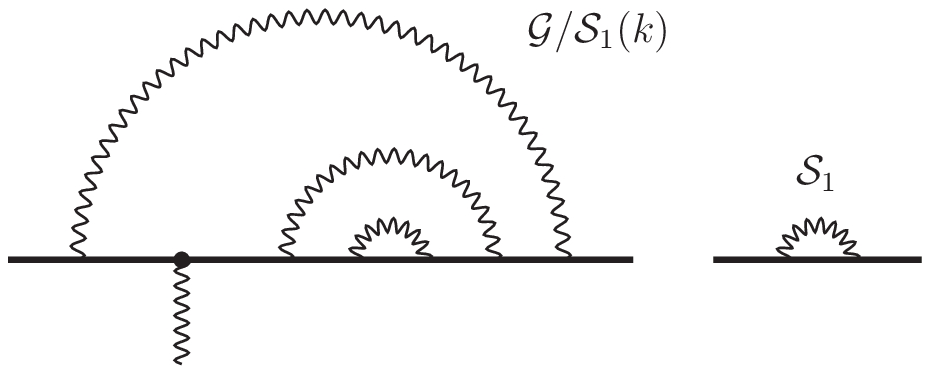}
\item $R_{46}$ \\
\includegraphics[scale=0.9]{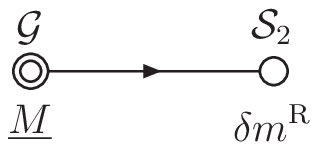} \hskip8em
\includegraphics[scale=0.6]{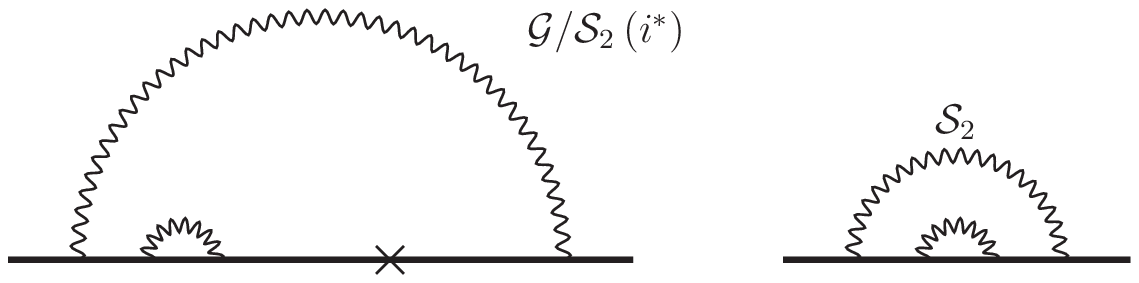}
\item $I_{1237}$ \\
\includegraphics[scale=0.9]{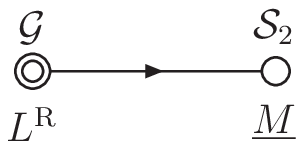} \hskip8em
\includegraphics[scale=0.6]{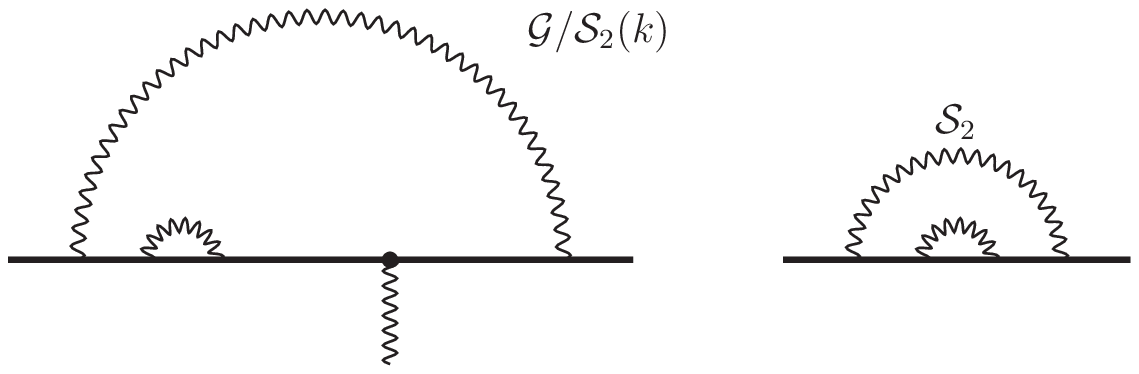}
\item $I_{123467}$ \\
\includegraphics[scale=0.9]{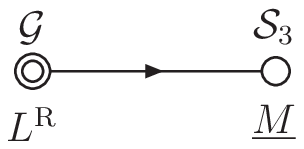} \hskip8em
\includegraphics[scale=0.6]{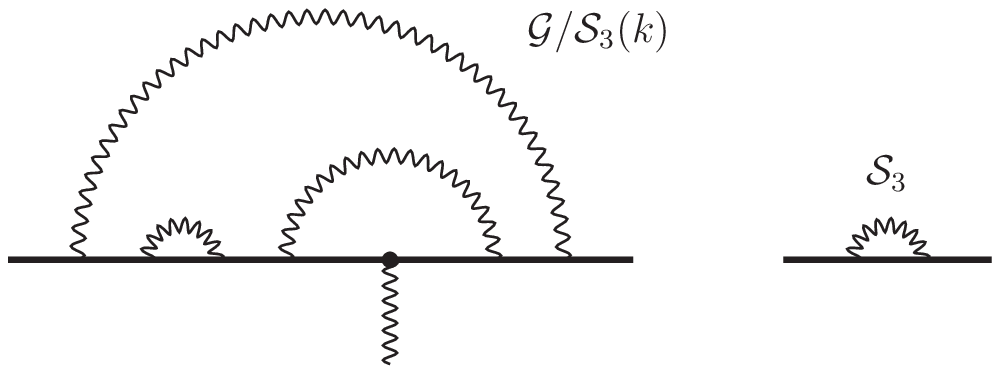}
\item $I_{137} R_{46}$ \\
\includegraphics[scale=0.9]{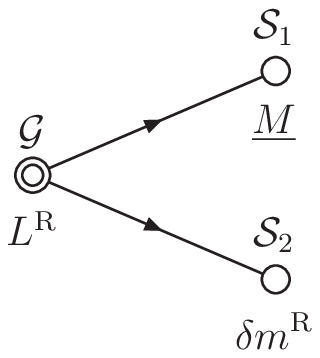} \hskip8em
\includegraphics[scale=0.6]{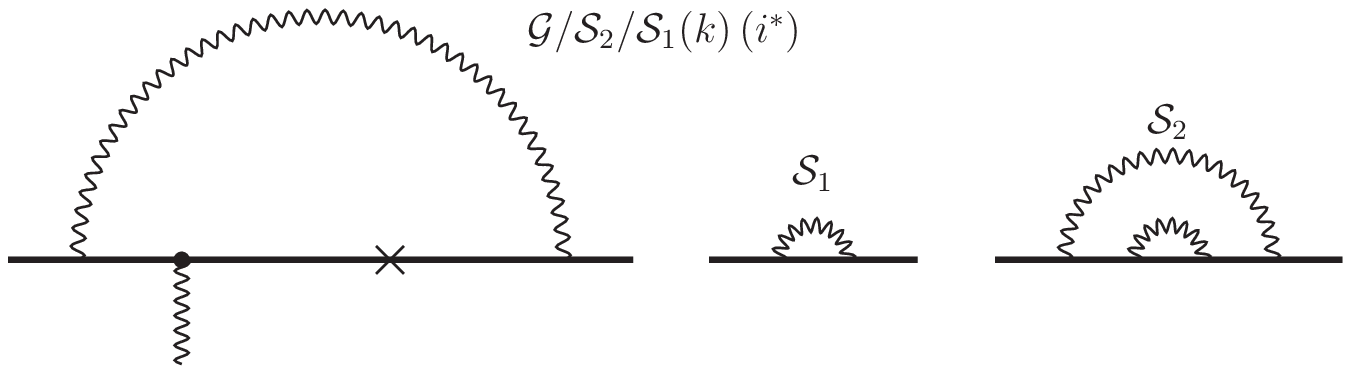}
\item $I_{1237} I_{46}$ \\
\includegraphics[scale=0.9]{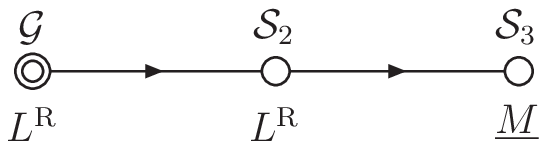} \hskip3em
\includegraphics[scale=0.6]{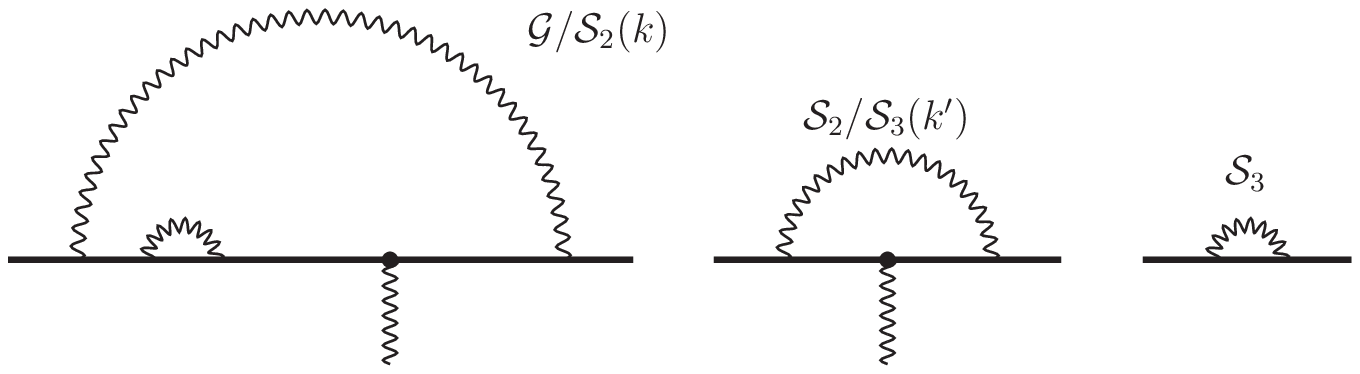}
\end{itemize}



\bibliographystyle{apsrev}
\bibliography{b}

\begin{thebibliography}{18}
\expandafter\ifx\csname natexlab\endcsname\relax\def\natexlab#1{#1}\fi
\expandafter\ifx\csname bibnamefont\endcsname\relax
  \def\bibnamefont#1{#1}\fi
\expandafter\ifx\csname bibfnamefont\endcsname\relax
  \def\bibfnamefont#1{#1}\fi
\expandafter\ifx\csname citenamefont\endcsname\relax
  \def\citenamefont#1{#1}\fi
\expandafter\ifx\csname url\endcsname\relax
  \def\url#1{\texttt{#1}}\fi
\expandafter\ifx\csname urlprefix\endcsname\relax\def\urlprefix{URL }\fi
\providecommand{\bibinfo}[2]{#2}
\providecommand{\eprint}[2][]{\url{#2}}

\bibitem[{\citenamefont{Kusch and Foley}(1947)}]{Kusch:1947}
\bibinfo{author}{\bibfnamefont{P.}~\bibnamefont{Kusch}} \bibnamefont{and}
  \bibinfo{author}{\bibfnamefont{H.~M.} \bibnamefont{Foley}},
  \bibinfo{journal}{Phys. Rev.} \textbf{\bibinfo{volume}{72}},
  \bibinfo{pages}{1256} (\bibinfo{year}{1947}).

\bibitem[{\citenamefont{Van~Dyck et~al.}(1987)\citenamefont{Van~Dyck,
  Schwinberg, and Dehmelt}}]{VanDyck:1987ay}
\bibinfo{author}{\bibfnamefont{R.~S.} \bibnamefont{Van~Dyck}},
  \bibinfo{author}{\bibfnamefont{P.~B.} \bibnamefont{Schwinberg}},
  \bibnamefont{and} \bibinfo{author}{\bibfnamefont{H.~G.}
  \bibnamefont{Dehmelt}}, \bibinfo{journal}{Phys. Rev. Lett.}
  \textbf{\bibinfo{volume}{59}}, \bibinfo{pages}{26} (\bibinfo{year}{1987}).

\bibitem[{\citenamefont{Odom et~al.}(2006)\citenamefont{Odom, Hanneke, D'Urso,
  and Gabrielse}}]{Odom:2006gg}
\bibinfo{author}{\bibfnamefont{B.}~\bibnamefont{Odom}},
  \bibinfo{author}{\bibfnamefont{D.}~\bibnamefont{Hanneke}},
  \bibinfo{author}{\bibfnamefont{B.}~\bibnamefont{D'Urso}}, \bibnamefont{and}
  \bibinfo{author}{\bibfnamefont{G.}~\bibnamefont{Gabrielse}},
  \bibinfo{journal}{Phys. Rev. Lett.} \textbf{\bibinfo{volume}{97}},
  \bibinfo{pages}{030801} (\bibinfo{year}{2006}).

\bibitem[{\citenamefont{Gabrielse et~al.}(2007)\citenamefont{Gabrielse,
  Hanneke, Kinoshita, Nio, and Odom}}]{Gabrielse:2006ggE}
\bibinfo{author}{\bibfnamefont{G.}~\bibnamefont{Gabrielse}},
  \bibinfo{author}{\bibfnamefont{D.}~\bibnamefont{Hanneke}},
  \bibinfo{author}{\bibfnamefont{T.}~\bibnamefont{Kinoshita}},
  \bibinfo{author}{\bibfnamefont{M.}~\bibnamefont{Nio}}, \bibnamefont{and}
  \bibinfo{author}{\bibfnamefont{B.}~\bibnamefont{Odom}},
  \bibinfo{journal}{Phys. Rev. Lett.} \textbf{\bibinfo{volume}{98}},
  \bibinfo{pages}{039902(E)} (\bibinfo{year}{2007}).

\bibitem[{\citenamefont{Mohr and Taylor}(2005)}]{CODATA:2002}
\bibinfo{author}{\bibfnamefont{P.~J.} \bibnamefont{Mohr}} \bibnamefont{and}
  \bibinfo{author}{\bibfnamefont{B.~N.} \bibnamefont{Taylor}},
  \bibinfo{journal}{Rev. Mod. Phys.} \textbf{\bibinfo{volume}{77}},
  \bibinfo{pages}{1} (\bibinfo{year}{2005}).

\bibitem[{\citenamefont{Melnikov and Vainshtein}(2004)}]{Melnikov:2003xd}
\bibinfo{author}{\bibfnamefont{K.}~\bibnamefont{Melnikov}} \bibnamefont{and}
  \bibinfo{author}{\bibfnamefont{A.}~\bibnamefont{Vainshtein}},
  \bibinfo{journal}{Phys. Rev.} \textbf{\bibinfo{volume}{D70}},
  \bibinfo{pages}{113006} (\bibinfo{year}{2004}).

\bibitem[{\citenamefont{Gabrielse et~al.}(2006)\citenamefont{Gabrielse,
  Hanneke, Kinoshita, Nio, and Odom}}]{Gabrielse:2006gg}
\bibinfo{author}{\bibfnamefont{G.}~\bibnamefont{Gabrielse}},
  \bibinfo{author}{\bibfnamefont{D.}~\bibnamefont{Hanneke}},
  \bibinfo{author}{\bibfnamefont{T.}~\bibnamefont{Kinoshita}},
  \bibinfo{author}{\bibfnamefont{M.}~\bibnamefont{Nio}}, \bibnamefont{and}
  \bibinfo{author}{\bibfnamefont{B.}~\bibnamefont{Odom}},
  \bibinfo{journal}{Phys. Rev. Lett.} \textbf{\bibinfo{volume}{97}},
  \bibinfo{pages}{030802} (\bibinfo{year}{2006}).

\bibitem[{\citenamefont{Kinoshita and Nio}(2006)}]{Kinoshita:2005sm}
\bibinfo{author}{\bibfnamefont{T.}~\bibnamefont{Kinoshita}} \bibnamefont{and}
  \bibinfo{author}{\bibfnamefont{M.}~\bibnamefont{Nio}},
  \bibinfo{journal}{Phys. Rev. D} \textbf{\bibinfo{volume}{73}},
  \bibinfo{pages}{053007} (\bibinfo{year}{2006}).

\bibitem[{\citenamefont{Kinoshita}(1990)}]{Kinoshita:1990}
\bibinfo{author}{\bibfnamefont{T.}~\bibnamefont{Kinoshita}}, in
  \emph{\bibinfo{booktitle}{Quantum electrodynamics}}, edited by
  \bibinfo{editor}{\bibfnamefont{T.}~\bibnamefont{Kinoshita}}
  (\bibinfo{publisher}{World Scientific, Singapore}, \bibinfo{year}{1990}), pp.
  \bibinfo{pages}{218--321}, \bibinfo{note}{(Advanced series on directions in
  high energy physics, 7)}.

\bibitem[{\citenamefont{Aoyama et~al.}(2006)\citenamefont{Aoyama, Hayakawa,
  Kinoshita, and Nio}}]{Aoyama:2005kf}
\bibinfo{author}{\bibfnamefont{T.}~\bibnamefont{Aoyama}},
  \bibinfo{author}{\bibfnamefont{M.}~\bibnamefont{Hayakawa}},
  \bibinfo{author}{\bibfnamefont{T.}~\bibnamefont{Kinoshita}},
  \bibnamefont{and} \bibinfo{author}{\bibfnamefont{M.}~\bibnamefont{Nio}},
  \bibinfo{journal}{Nucl. Phys. B} \textbf{\bibinfo{volume}{740}},
  \bibinfo{pages}{138} (\bibinfo{year}{2006}).

\bibitem[{\citenamefont{Lepage}(1978)}]{vegas}
\bibinfo{author}{\bibfnamefont{G.~P.} \bibnamefont{Lepage}},
  \bibinfo{journal}{J. Comput. Phys.} \textbf{\bibinfo{volume}{27}},
  \bibinfo{pages}{192} (\bibinfo{year}{1978}).

\bibitem[{\citenamefont{Kinoshita and Lindquist}(1990)}]{Kinoshita:1981wm}
\bibinfo{author}{\bibfnamefont{T.}~\bibnamefont{Kinoshita}} \bibnamefont{and}
  \bibinfo{author}{\bibfnamefont{W.~B.} \bibnamefont{Lindquist}},
  \bibinfo{journal}{Phys. Rev. D} \textbf{\bibinfo{volume}{42}},
  \bibinfo{pages}{636} (\bibinfo{year}{1990}).

\bibitem[{\citenamefont{Cvitanovi\'c and
  Kinoshita}(1974{\natexlab{a}})}]{Cvitanovic:1974uf}
\bibinfo{author}{\bibfnamefont{P.}~\bibnamefont{Cvitanovi\'c}}
  \bibnamefont{and}
  \bibinfo{author}{\bibfnamefont{T.}~\bibnamefont{Kinoshita}},
  \bibinfo{journal}{Phys. Rev. D} \textbf{\bibinfo{volume}{10}},
  \bibinfo{pages}{3978} (\bibinfo{year}{1974}{\natexlab{a}}).

\bibitem[{\citenamefont{Cvitanovi\'c and
  Kinoshita}(1974{\natexlab{b}})}]{Cvitanovic:1974sv}
\bibinfo{author}{\bibfnamefont{P.}~\bibnamefont{Cvitanovi\'c}}
  \bibnamefont{and}
  \bibinfo{author}{\bibfnamefont{T.}~\bibnamefont{Kinoshita}},
  \bibinfo{journal}{Phys. Rev. D} \textbf{\bibinfo{volume}{10}},
  \bibinfo{pages}{3991} (\bibinfo{year}{1974}{\natexlab{b}}).

\bibitem[{\citenamefont{Vermaseren}(2000)}]{Vermaseren:2000nd}
\bibinfo{author}{\bibfnamefont{J.~A.~M.} \bibnamefont{Vermaseren}}
  (\bibinfo{year}{2000}), \eprint{math-ph/0010025}.

\bibitem[{\citenamefont{Laporta and Remiddi}(1996)}]{Laporta:1996mq}
\bibinfo{author}{\bibfnamefont{S.}~\bibnamefont{Laporta}} \bibnamefont{and}
  \bibinfo{author}{\bibfnamefont{E.}~\bibnamefont{Remiddi}},
  \bibinfo{journal}{Phys. Lett.} \textbf{\bibinfo{volume}{B379}},
  \bibinfo{pages}{283} (\bibinfo{year}{1996}).

\bibitem[{\citenamefont{Aoyama et~al.}(2007)\citenamefont{Aoyama, Hayakawa,
  Kinoshita, and Nio}}]{Aoyama:2007dv}
\bibinfo{author}{\bibfnamefont{T.}~\bibnamefont{Aoyama}},
  \bibinfo{author}{\bibfnamefont{M.}~\bibnamefont{Hayakawa}},
  \bibinfo{author}{\bibfnamefont{T.}~\bibnamefont{Kinoshita}},
  \bibnamefont{and} \bibinfo{author}{\bibfnamefont{M.}~\bibnamefont{Nio}}
  (\bibinfo{year}{2007}), \eprint{arXiv:0706.3496 [hep-ph]}.

\bibitem[{\citenamefont{Kinoshita and Nio}(2004)}]{Kinoshita:2004wi}
\bibinfo{author}{\bibfnamefont{T.}~\bibnamefont{Kinoshita}} \bibnamefont{and}
  \bibinfo{author}{\bibfnamefont{M.}~\bibnamefont{Nio}},
  \bibinfo{journal}{Phys. Rev.} \textbf{\bibinfo{volume}{D70}},
  \bibinfo{pages}{113001} (\bibinfo{year}{2004}).

\end{thebibliography}

\end{document}